# Silicon carbide for integrated photonics

*Ailun Yi[1,3,#], Chengli Wang[1,2,#], Liping Zhou[1,2], Yifan Zhu[1], Min Zhou[1], Shibin Zhang[1,3], Tiangui You[1,2], Jiaxiang Zhang[1,2,a)], and Xin Ou[1,2,b)]*

[1] State Key Laboratory of Functional Materials for Informatics, Shanghai Institute of Microsystem and Information Technology, Chinese Academy of Sciences, Shanghai, 200050, China

[2] The Center of Materials Science and Optoelectronics Engineering, University of Chinese Academy of Sciences, Beijing 100049, China

[3] XOI Technology Co., Ltd, Shanghai 201899, China

# These authors contributed equally to this work

Corresponding author email:

[a)] jiaxiang.zhang@mail.sim.ac.cn; [b)] ouxin@mail.sim.ac.cn

**Abstract**

Photonic integrated circuits (PICs) based on lithographically patterned waveguides provide a scalable approach for manipulating photonic bits, enabling seminal demonstrations of a wide range of photonic technologies with desired complexity and stability. While the next generation of applications such as ultra-high speed optical transceivers, neuromorphic computing and terabit-scale communications demand further lower power consumption and higher operating frequency. Complementing the leading silicon-based material platforms, the third-generation semiconductor, silicon carbide (SiC), offers a significant opportunity towards the advanced development of PICs in terms of its broadest range of functionalities, including wide bandgap, high optical nonlinearities, high refractive index, controllable artificial spin defects and CMOS-compatible fabrication process. The superior properties of SiC have enabled a plethora of nano-photonic explorations, such as waveguides, micro-cavities, nonlinear frequency converters and optically-active spin defects. These remarkable progresses have prompted the rapid development of advanced SiC PICs for both classical and quantum applications. Here, we provide an overview of SiC-based integrated photonics, presenting the latest progress on investigating its basic optoelectronic properties, as well as the recent developments in the fabrication of several typical approaches for light confinement structures that form the basic building blocks for low-loss, multi-functional and industry-compatible integrated photonic platform. Moreover, recent works employing SiC as optically-readable spin hosts for quantum information applications are also summarized and highlighted. As a still-developing integrated photonic platform, prospects and challenges of utilizing SiC material platforms in the field of integrated photonics are also discussed.

# Contents

## 1. Introduction

Photons are widely accepted as information carriers due to their prominent characteristics, such as speed-of-light transmission, weak interaction, high bandwidth, low propagation loss and power consumption. In analogous to integrated microelectronic circuits, confining photons in nano-photonic structures enables the miniaturization of conventional optical components to realize PICs, thus offering an opportunity to implement photonic technologies on a single chip with desired scalability and stability. Moreover, the fundamental interest in PICs lies in its inheritance of microelectronic structures with a small footprint and high integration density. These genuine features have stimulated a wide range of photonic applications from daily life to the most advanced science, such as telecommunications[1], biophotonics[2,3] and quantum networks[4].

The advanced development of PICs calls for suitable material platforms that are capable of confining light in nanometer-scale. At present, silicon represents the core material for PICs owing the mature CMOS-compatible device fabrication processes[5] which allow for manufacturing large-scale PICs with a little increase of cost. Benefiting from the successful commercialization of silicon-on-insulator (SOI), silicon-based PICs have attracted increasing attention in the past two decades due to their widespread applications in information science and technology. Nowadays, it has evolved to a leading platform to perform a wide variety of photonic applications[6]. Despite the remarkable success, the absence of light sources due to the indirect band gap feature of silicon places a constraint on developing a truly integrated silicon-based optoelectronics platform. Although several ways of exploiting nonlinear light-matter interactions to coax silicon into optical functionality[6-8] have been reported, these effects turn out to be relatively weak. Moreover, the narrow indirect bandgap of silicon results in the cut-off wavelength at 1.1 μm, and the additional two-photon absorption (TPA) effect leads to a high loss at the telecom wavelengths[9]. Furthermore, the lack of the Pockels effect due to its centrosymmetric lattice structure makes silicon challenge to achieve desired light modulation speed and efficiency[6]. Towards the advanced integrated photonics with monolithically integrated light sources and efficient light modulators, exploring novel materials with ultra-low loss and high electro-optic coefficients is highly demanded.

Thus far, a variety of material platforms, such as III-V semiconducting compounds[10,11], lithium niobate ($LiNbO_3$, LN)[12,13], silicon nitride ($Si_3N_4$)[14], indium phosphide (InP)[15] and diamond[16,17] have been extensively investigated for PICs. Table 1 lists the key specifications of these materials. Among them, III-V compounds such as gallium arsenide (GaAs) are appealing as they have high refractive index, large second-order and Kerr nonlinearities[18]. Of interesting is that III-V compounds are direct semiconductors and they can be easily engineered for efficient on-chip laser sources. Despite the noteworthy potential, III-V semiconducting materials are prone to a relatively larger propagation loss which makes the improvement of the circuit complexity of III-V photonics cumbersome. Another platform, $LiNbO_3$, known as "silicon of photonics", has recently been considered as one of the most promising platforms for PICs[19]. It simultaneously possesses larger transparent window range (0.35-5.5 μm), the excellent second-order ($r_{33}=2.52\times10^{-11}$ m $V^{-1}$) and third-order ($1.8\times10^{-6}$ $cm^2$ $GW^{-1}$) nonlinearities, ultra-low optical losses[20] and commercially wafer-scale lithium niobate-on-insulator (LNOI) substrates. With the recent breakthrough in dry-etching techniques, integrated photonics based on LNOI have made significant progress, with tremendous applications in ultrafast light modulation and photonic switching. However, scaling beyond the efficient light modulations is presently limited by relatively large losses from the use of off-chip laser sources. Although the light sources can be integrated using hybrid approaches[21], the low efficiency and sophisticated integrated methods render practical applications inconvenient. Furthermore, its incompatibility with the CMOS fabrication methods makes LNOI cumbersome to realize low-cost and high-yield PICs.

**Table 1**. Key properties of dominant material platforms for PICs

| | Si | GaAs | LN | SiC[(a)] | Diamond |
|---|---|---|---|---|---|
| Bandgap [eV] | 1.1 | 1.42 | 4 | 2.3-3.2 | 5.5 |
| Transparent window [um] | 1.1-5.5 | 0.9-17.3 | 0.35-5.5 | 0.37-5.6 | 0.22-20 |
| Refractive index | 3.5 | 3.4 | 2.2 | 2.5-2.7 | 2.4 |
| Thermal conductivity [ W $cm^{-1}$ $K^{-1}$] | 1.1 | 0.55 | 0.042 | 3.6-4.9 | 10-25 |
| Mohs Hardness [1-10] | 7 | 5 | 5 | 9.2-9.5 | 10 |
| $Q\times f$ (GHz)[22] | 15.4 | 40.3 | --[(b)] | 30,000 | 3.7 |
| Second-order nonlinearities [$d_{ij}$, pm $V^{-1}$] | -- | $d_{14}$=105 (@1.5 μm) | $d_{33}$=25.2 (@1.06 μm) | $d_{33}$=-13~-24 (@1.06 μm) | -- |
| Electro-optic coefficient (pm/V) | Carrier plasma effect | $r_{14}$=1.2, Pockels effect | $r_{33}$=30.8, Pockels effect | $r_{33}$=1.8, Pockels effect | -- |
| Third-order nonlinearities [$\times10^{-7}$ $cm^2$ $GW^{-1}$] | 450 (@1.55 μm)[23] | 1590 (@1.55 μm)[24] | 18 (@1.06 μm)[25] | 60-80 (@1.55 μm)[26] | 8.2 (@1.55 μm)[17] |
| Waveguide losses | Moderate 1 dB $cm^{-1}$[(c)] | Moderate 0.4 dB $cm^{-1}$ [(d)] | Ultra-low 0.027 dB $cm^{-1}$ | Moderate 0.15 dB $cm^{-1}$ [(e)] | Moderate 0.34 dB $cm^{-1}$ [17] |
| Nonclassical photon emission[21] | Probabilistic $\chi^{(3)}$, RT | On-demand QDs, low temperature | Probabilistic $\chi^{(2)}$, RT | On-demand color centers, RT | On-demand color centers, RT |
| Commercialized Wafer | Yes, 1960's | Yes, 1980's | Yes, 1990's | Yes, 1990's | -- |
| Thin film-On-insulator | Yes, 2000's | Yes | Yes, 2010's | Yes | -- |

[a)] SiC includes many polytypes, and their properties vary slightly. [b)] '--' indicates no functionality reported; [c)] The data is quoted from commercial silicon photonics foundry[27]; [d)] The state-of-the-art waveguide propagation loss realized in (Al)GaAs on insulator platform[24,28]; e) Calculated value, not shown in the article[29].

To circumvent the challenges in sources and fast light modulators, tremendous efforts have been devoted to explore wide-band semiconductors for PICs, with representative examples being diamond and silicon carbide (SiC)[17,30-33]. Optically, diamond has a wide band gap of around 5.5 eV and thus the diamond is transparent from the UV to IR. In addition, it has large Raman gain and a high refractive index (n = 2.4) which is intensity-dependent. All these material properties make diamond a promising choice for PICs. Yet diamond is not mature for standard device fabrication protocols, and its full potential cannot be realized unless the technology to fabricate the thin film diamond is maturely solved. On the contrary, the third-generation semiconductor, *i.e.*, SiC, is a well-known CMOS-compatible material, and it has been broadly explored for electronic devices by exploiting the well-developed device processing protocols. Moreover, SiC provides many properties suitable for PICs, such as wide transparent window from the ultraviolet to the mid-infrared[34], the relatively high second-order ($d_{33}$ = -13~-24 pm $V^{-1}$)[35], third-order nonlinearity (~$10^{-6}$ $cm^2$ $GW^{-1}$)[36] and high refractive index of about 2.6 at the telecom wavelengths. Inspired by these outstanding optical properties, SiC has so far attracted a great deal of research attention. The recent successful realization of various nano-photonic devices highlights its great suitability for next-generation integrated photonics[30,37,38]. Another major advantage of SiC is the flexibility to incorporate various "artificial defects". More recently, SiC has been investigated to host deep optically active defects so that advanced quantum devices have been successfully demonstrated[32,39,40]. When considering the special capability of hosting active spin-defects, together with the above wide optical transparent window, high refractive index and compatibility with CMOS fabrication methods, SiC may soon enjoy a renaissance in photonics and become a competing material platform for PICs.

At present, the most extensively studied polytypes of SiC include 4H-SiC and 6H-SiC with hexagonal lattice structures and 3C-SiC with a zinc-blende crystal structure (cubic). The reliable fabrication of these different polytypes has promoted the development of various elementary nano-photonic devices crucial for densely integrated high-performance photonic integrated circuits. Researchers devoted primitive exploration on the 3C-SiC platform due to its relatively simple epitaxial process on Si[41-43]. Additionally, significant progress has been made recently in photonic cavities on epitaxially grown 4H-SiC[44] and micro-ring resonators in 4H-SiC-on-insulator (4H-SiCOI) prepared by bonding, thinning and polishing techniques[30,32,45]. The recent successful demonstration of the coupling of single silicon vacancy defect-based light emitters with nano-photonic cavities in 4H-SiC-on-insulator substrate represents a hallmark work toward integrated quantum and nonlinear photonics[32,39]. Although SiC as a host for integrated photonics and novel quantum systems is still in its infancy, the rapid progress made in recent years indicates it could be one of the most prominent monolithic platforms to be investigated or employed in the future.

Given the superior optical properties and the ability to incorporate optically active defects for quantum applications, SiC is able to serve as a single material platform to develop linear and nonlinear optical systems, novel quantum devices and possibly integrated quantum photonic circuits. In this article, we review the recent progress on the development of various SiC photonic platforms and their relevant nano-photonic devices. In addition, we provide a comprehensive summary on works that explored artificial defects which are utilized as single photon emitters for both spintronic applications and quantum light emitters. The article is structured as follows. Chapter 2 summarizes the fabrication of light confinement structures based on SiC platforms, including selective etching, angle etching, dopant-selective etching and a particular focus on SiC-on-insulator (SiCOI) structures. In chapter 3, we review the recent development of nano-photonic devices on SiC platforms, including optical resonators and waveguides. Chapter 4 reviews the current status of SiC in nonlinear optics. In chapter 5, the optomechanical properties of SiC are introduced. Chapter 6 summarizes the key results of color center-based defects explored in SiC and covers the recent progress in the integration of these color centers with photonic structures. Finally, the prospects and challenges of SiC in the foreseen development of large-scale PICs, as well as the relevant applications in nonlinear and quantum photonics are discussed.

## 2. Fabrication of light confinement structures based on SiC platforms

SiC is known as a wide bandgap semiconductor with abundant polytypes. So far, there exist 250 different polytypes but only 3C ($\bar{4}3m$), 4H ($6mm$) and 6H ($6mm$)-SiC can be stably grown and commercially available for the use of optoelectronic devices. Pioneering photonics research mainly focused on bulk SiC[46], hetero-epitaxial 3C-SiC[47,48], homo-epitaxial 4H-SiC thin film[49], deposited amorphous SiC ($a$-SiC) on Si[50] and SiCOI[51]. Before the development of functional photonic devices on kinds of SiC platforms, the light confinement structures need to be set up as the fundamental parts of PICs. This chapter summarizes a series of classical light confinement structures and their fabrication processes in SiC photonics as the first main part of this review.

### 2.1 Selective etching for 3C-silicon carbide on silicon

3C-SiC has been the primary choice for fabricating optoelectronic devices due to the easy thin-film growth on silicon substrate by using a hetero-epitaxial method. The successful preparation of 3C-SiC on a Si wafer was reported in 1959[52]. It is well-known that silicon wafers can be produced on a very large scale, with good quality and relatively low price. Consequently, the hetero-epitaxy of 3C-SiC on Si (epi-3C-SiC) can provide a cost-effective single-crystal SiC film on very large silicon wafers. From the photonic applications point of view, integration of 3C-SiC on Si wafers could bring several important advantages[50,53-56]. First, both are compatible with the CMOS fabrication methods, which offers the possibility to manufacture large-scale photonic circuits with high efficiency and low cost. Second, silicon facilitates an integration with electronic circuits which is desirable for real-world photonic applications. Light confinement structures were generally obtained by undercutting the beneath substrate, as shown in Fig. 1(a) and (b). With the undercutting of Si, the light could be confined in the SiC layer by the low refractive index of air.

However, based on this configuration, the photonic structures commonly require sophisticated processing flow, which limits the device functionality and versatility in integration with other materials. The low yield and poor reliability further make the fabrication of scalable photonic devices inconvenient, such as micro-ring structure, etc[48,54,57]. In addition, growth of 3C-SiC on silicon substrate by CVD technique also comes with a severe limitation from the large lattice mismatch of ~20% between Si and 3C-SiC, as well as the ~8% thermal expansion mismatch[58]. Such limitation gives rise to unavoidable stacking faults and other extended defects in the grown 3C-SiC layer. The growth of 3C-SiC on Si also induces defects in the silicon at the interface, such as voids and pits going down deep into the Si substrate, because of the out-diffusion of Si from the substrate and its incorporation into the SiC growth process. Thus, with the development of the SiC photonics field, the selective etching method and the platform of 3C-SiC on Si gradually fall into disuse within recent works.

### 2.2 Angled etching on silicon carbide bulk

Compared to epi-3C-SiC/Si, 4H-SiC shows both better performance in terms of industry maturity and the capability to host plenty of spin-defects for quantum applications. In this context, 4H-SiC is attracting increasing research interest in nonlinear photonics[32,59] and quantum photonics[39,40,59] rather than 3C[57] or 6H[60]. However, unlike epi-3C-SiC/Si, the lack of hetero-epitaxial growth and scalable fabrication techniques becomes the major factor which prevents more wide-spread development and application of 4H-SiC photonics. Similar to the fabrication of photonic cavities in diamond[61], angled etching technique provides a viable route to fabricate nano-photonic structures on bulk 4H-SiC platform. Recently, nano-photonic waveguides were reported by C. Babin *et al.*[39]. The detailed fabrication method is shown in Fig. 1(c). First, vertical volume structures were fabricated by plasma dry etching with a patterned mask on bulk 4H-SiC. Then, angled plasma was used to realize suspended nanobeam structures. In this step, ion sheath control plate or Faraday Cage was used to simultaneously form angled sidewall on both sides[39]. The angled etching method always creates triangle waveguides with highly efficient light confinement by the low refractive index of the surrounding air. Figure. 1(d) shows a typical scanning electron beam microscopic (SEM) image of the fabricated suspended waveguides. Thereby a simple and effective fabrication process is supported by angled etching for SiC bulk photonics, especially for quantum physics research[39]. This method based on two-stage dry-etching process has been demonstrated to keep the spin optical properties of SiC. However, the suspended device structures are always limited into nanobeam and crystal cavities[39,62,63]. A system-level integrated photonic devices is still a challenge to be realized on SiC bulk.

### 2.3 Dopant-selective etching on multiple 4H-SiC layers

Another approach to realize photonic structures on bulk SiC substrate is to employ a dopant-dependent selective etching technique, followed by a selective etching[44,64-66]. Figure. 1(e) depicts the details of this fabrication method. The *p*-doped SiC layer was epitaxially grown on a *n*-type bulk SiC firstly and then the photonic structures were patterned and developed successively by electron beam lithography and dry etching techniques. Dopant-selective photoelectrochemical etching was adapted, which selectively etched the *n*-type substrate beneath the patterned devices whilst keeping *p*- type 4H-SiC unchanged. Vice versa, the *p*-type etching could also be realized[44]. With this process, photonic crystal nanobeam cavities were fabricated, as shown by the SEM image. The dopant-selective etching method provides an alternative means to fabricate nano-photonic structures, including nano-photonic waveguides and crystal cavities[65].

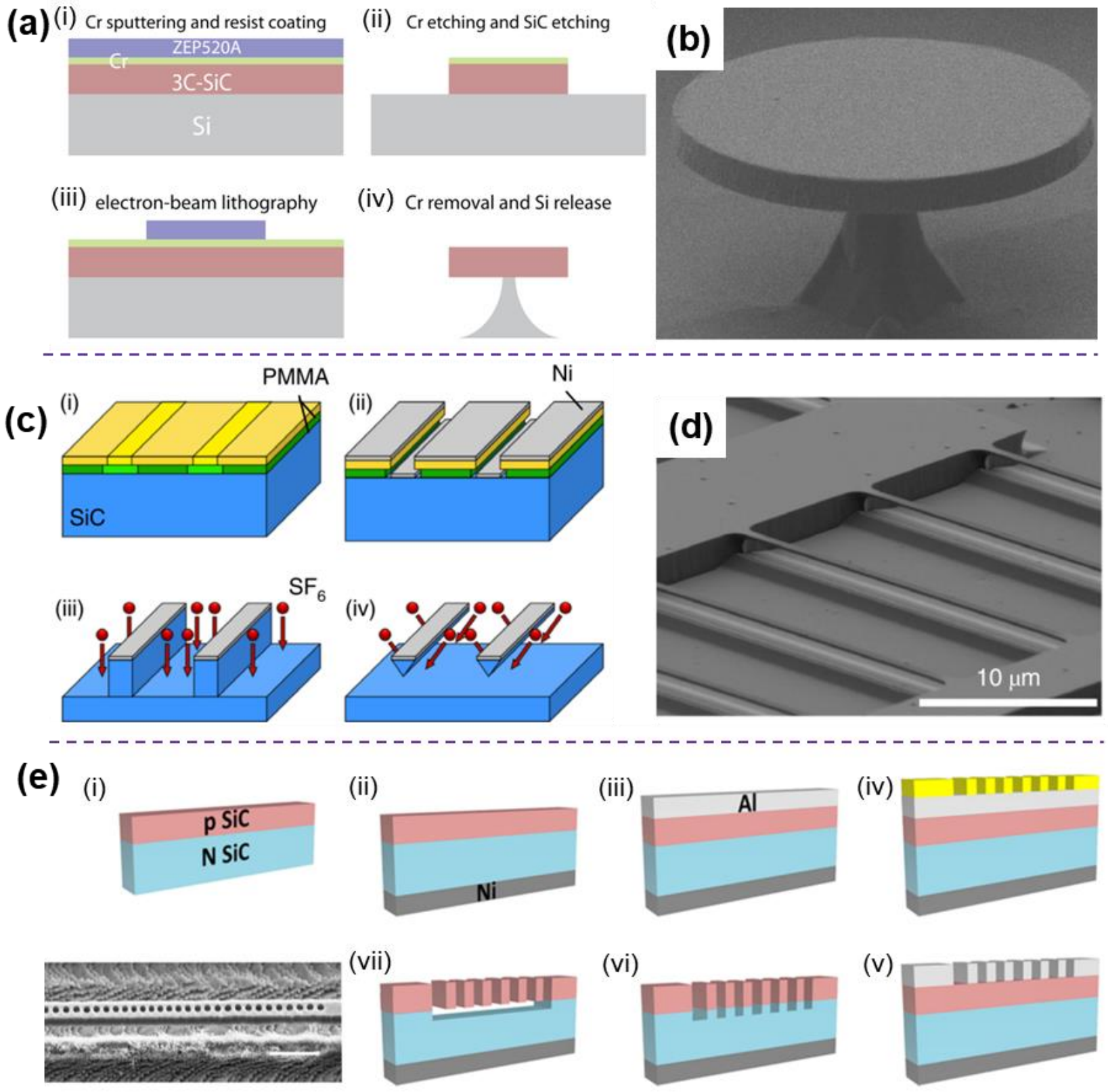

**Figure 1.** Different methods for fabricating nano-photonic devices on bulk SiC substrate. (a) Selective etching of 3C-SiC on Si. (b) SEM image of a fabricated micro-disk on 3C-SiC on Si. Reprinted with the permission from X. Lu *et al.,* Appl. Phys. Lett. **104**, 181103 (2014). Copyright 2014 AIP Publishing LLC[48]. (c) Processing flow of angled etching on bulk 4H-SiC substrate. (i) Deposition of a double layer of PMMA mask. (ii) Deposition of a nickel mask. (iii) A straight reactive ion etching based on $SF_6$ plasma creates deep trenches. (iv) An angled $SF_6$ plasma etch created triangular cross-section waveguides. (d) SEM image of an array of the fabricated photonic crystal nanobeams. Reprinted with the permission from C. Babin *et al.,* Nat. Mater. **21**, 67 (2022). Copyright 2021 Springer Nature Limited[39]. (e) Dopant-dependent selective etching on bulk 4H-SiC. (i) The starting material was an epilayer of *p*-type 4H-SiC homoepitaxially grown on a substrate of *n*-type 4H-SiC. (ii) Ni layer was thermally evaporated on the backside of each sample. (iii) Al layer was thermally evaporated to serve as an etch mask. (iv) Electron beam lithography defines the desired pattern in PMMA resist. (v) The pattern was transferred to Al by reactive ion etching (RIE) with a chlorine-based chemistry. (vi) The pattern was transferred to SiC by a further fluorine-based RIE step. (vii) Dopant-selective photoelectrochemical etching was used to remove the *n*-type substrate beneath the patterned devices. Reprinted with the permission from D.O. Bracher *et al.,* Nano Lett. **15**, 6202 (2015). Copyright 2015 American Chemical Society[64].

### 2.4 Silicon carbide-on-insulator (SiCOI)

With the previous discussion, we suggest that the efficient, stable and generic structure is still extremely necessary for the realization of integrated photonics system on SiC platforms rather than individual devices. A long-proposed solution to alleviate this complication is to use SiC thin film on insulator, so called SiC-on-insulator (SiCOI). Akin to SOI, SiCOI ensures a relatively large refractive index contrast in the vertical direction. Thereby waveguides and resonators can be efficiently patterned on the SiC layer to confine light, which is a critical requirement for scalable integrated photonic architectures. In the last decade, active researches have been carried out to develop high quality SiCOI platforms[32,43,67].

#### 2.4.1 Optimized solution for 3C-SiCOI with epitaxial growth

As mentioned in section 2.1, epi-3C-SiC/Si shows an inappropriate structure and quality for current SiC photonics development. To overcome these complications, by borrowing the idea of direct wafer bonding technique from silicon photonics, T. Fan *et al.* reported an optimal process to fabricate a 3C-SiC on $SiO_2$ as shown in Fig. 2[43]. This structure ensures a high refractive index contrast for optical mode confinement in the 3C-SiC layer and isolation from the Si substrate because of the refractive indices of 3C-SiC and $SiO_2$ (2.6 and 1.44, respectively). Most importantly, the 3C-SiCOI platform is compatible with large-scale integration and suitable for the fabrication of SiC devices for linear, nonlinear, active and quantum optical applications.

To experimentally fabricate the 3C-SiCOI material, two Si dies were firstly prepared [Fig. 2(a), (d)], followed by epitaxially growing a 3C-SiC thin film on a Si die. Then the 3C-SiC thin-film was thinned down to 800 nm by a chemical mechanical polishing (CMP) process [Fig.2 (b)]. A $SiO_2$ layer with a thickness of 30 nm was deposited on the polished SiC film using atomic layer deposition (ALD) in order to provide a hydrophilic bonding interface with lower roughness and higher bonding strength. Thereafter, a 4 μm thick $SiO_2$ layer was grown on the other Si substrate (piece 2) by the wet oxidation process [Fig. 2(e)]. Considering the different thermal expansion coefficients between SiC and $SiO_2$/Si, the two pieces were then bonded into a monolithic piece with a wafer-scale hydrophilic bonding process at low temperature (300 °C) [Fig. 2(f)]. As the last step, the top Si substrate was removed and this resulted in the formation of a 3C-SiCOI substrate. The photo inset Fig. 2 shows a photographic image of "2×2 $cm^2$" die, and the complete cover of 3C-SiC indicates a nearly 100% yield. The same process without any changes could also be extended to develop wafer-scale SiCOI platforms for large-size (4, 6-inch) wafers. The quality of 3C-SiCOI was characterized by tunneling electron beam microscopy (TEM) as shown in Fig. 2(h)-(j). Fig. 2(h) shows the zoomed-out TEM image of the cross-sectional image of the 3C-SiCOI sample. It could be clearly observed that the high density of the diffraction pattern (line-shaped) near the top region of SiC film is due to the epitaxial lattice defect in this region. Figure. 2(i) shows the zoomed-in TEM image of the top layer of SiC film (region A) at an atomic scale. The high density of the lattice dislocation and the stacking faults, which lead to the optical loss, were clearly shown. Compared to region A, the other side of SiC film (region B) shown in Fig. 2(j) illustrates the uniformly stacked atoms and high crystalline quality of SiC. After removing the top layer by CMP process, the relatively high optical quality of 3C-SiC film remained.

The optimal 3C-SiCOI fabrication process leads to a significant improvement in SiC-based optical devices compared to the epi-3C-SiC/Si, also combining the potential for large wafer-scale integration. T. Fan *et al.* also first demonstrated an electro-optical micro-resonator device with a $Q$ factor up to $3 \times 10^4$ [38]. However, photonic micro-cavities fabricated on 3C-SiCOI were measured to have an upper limit of about $2 \times 10^5$. Indeed, this limitation stems from the intrinsic loss due to lattice dislocation of the 3C-SiC film. In order to improve the performance of optical devices on 3C-SiCOI platform, quality of the epitaxially-grown 3C-SiC thin film is amenable for further optimization. It should be noted that by using a buffer layer in combination with a proper adjustment of single-source precursors, higher quality of 3C-SiC thin film can be expected[58].

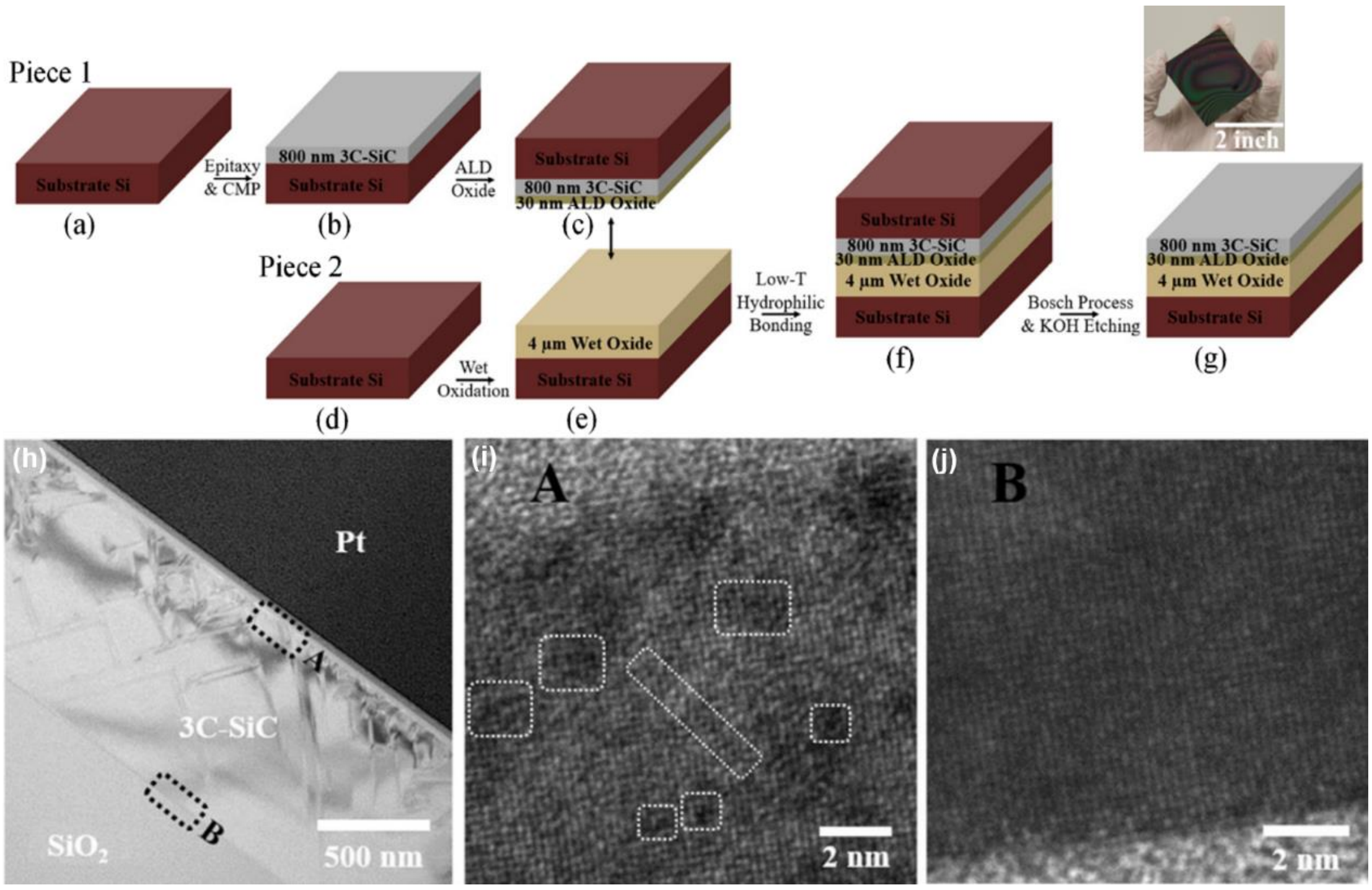

**Figure 2.** Fabrication process flow of the 3C-SiCOI material platform. (a) Piece 1, a prime Si wafer. (b) Epitaxial growth of 3C-SiC and surface smoothing by CMP. (c) Deposition of a 30 nm $SiO_2$ layer using ALD. (d) Piece 2, a prime Si wafer. (e) Growth of 4 µm $SiO_2$ by wet oxidation. (f) Piece 1 and piece 2 were bonded using a low-temperature hydrophilic bonding process. (g) Removal of the Si handle layer using Bosch process and KOH wet etching. Inset: a photo of the bonded SiCOI piece. (h) A zoomed-out TEM image of the SiCOI lamella. The line-shape patterns shown in the region where SiC exists were from the diffracted electrons due to the defects in the SiC lattice. The density of defects is reducing in the direction of SiC growth (from region A to B). (i) A zoomed-in TEM image of the top surface of the SiCOI lamella (region A in (a), also known as the transition layer). (j) A zoomed-in TEM image of the bottom layer of the SiCOI lamella (region B in (h)) where there is low density of defects in SiC. Reprinted with the permission from T. Fan *et al.,* Opt. Express **26**, 25814 (2018).

Copyright 2018 Optical Society of America[43].

### 2.4.2. 4H-SiCOI prepared by the ion-cutting technique

Although 3C-SiCOI material platform has shown a noteworthy potential for the development of PICs, the omnipresent material imperfections and stacking defects due to the lattice mismatch give rise to a large intrinsic absorption loss, leading to an upper bound to the performance of nano-photonic devices. On the contrary, hexagonal 4H-SiCOI has much lower intrinsic loss and it contributes a versatile material platform to implement photonic technologies. The fundamental reason of the interest in 4H-SiC stems from the well-developed technique for growing wafer-scale crystal with well-controlled crystalline quality[68]. Actually, wafer-scale growth and processing of 4H-SiC was developed in the 1990s for the applications in high-power electronics. Soon after, 4H- SiCOI were demonstrated using the same ion-implantation method as SOI. Indeed, applying 4H-SiCOI prepared by the ion-cutting process is practically favorable. First, this material does not compromise the crystalline integrity of the photonic device layer, thus allowing for an improvement in quality of photonic devices. Second, the ion-cutting technique is capable of fabricating wafer-scale 4H-SiC film on $SiO_2$/Si substrate with well-controlled thickness uniformity and surface roughness, offering the advantages for future foundry-based devices fabrication.

In the past decade, tremendous efforts have been devoted to fabricate the 4H-SiCOI substrate by using the ion-cutting technique[69-73]. Several efforts have been made in the fabrication of wafer-scale 4H-SiCOI substrate[32,67,74,75]. Figure. 3 shows the typical process flow of wafer-scale 4H-SiCOI with the ion-cutting technique, which includes a high-fluence ion implantation within a crystalline 4H-SiC bulk wafer (Fig. 3a), wafer bonding, thermal annealing and final transfer of 4H-SiC thin film. Since the thickness of the 4H-SiC thin film depends on the depth of the implanted $H^+$ ions, the thickness for specific applications could be precisely controlled through modification of the ion fluence.

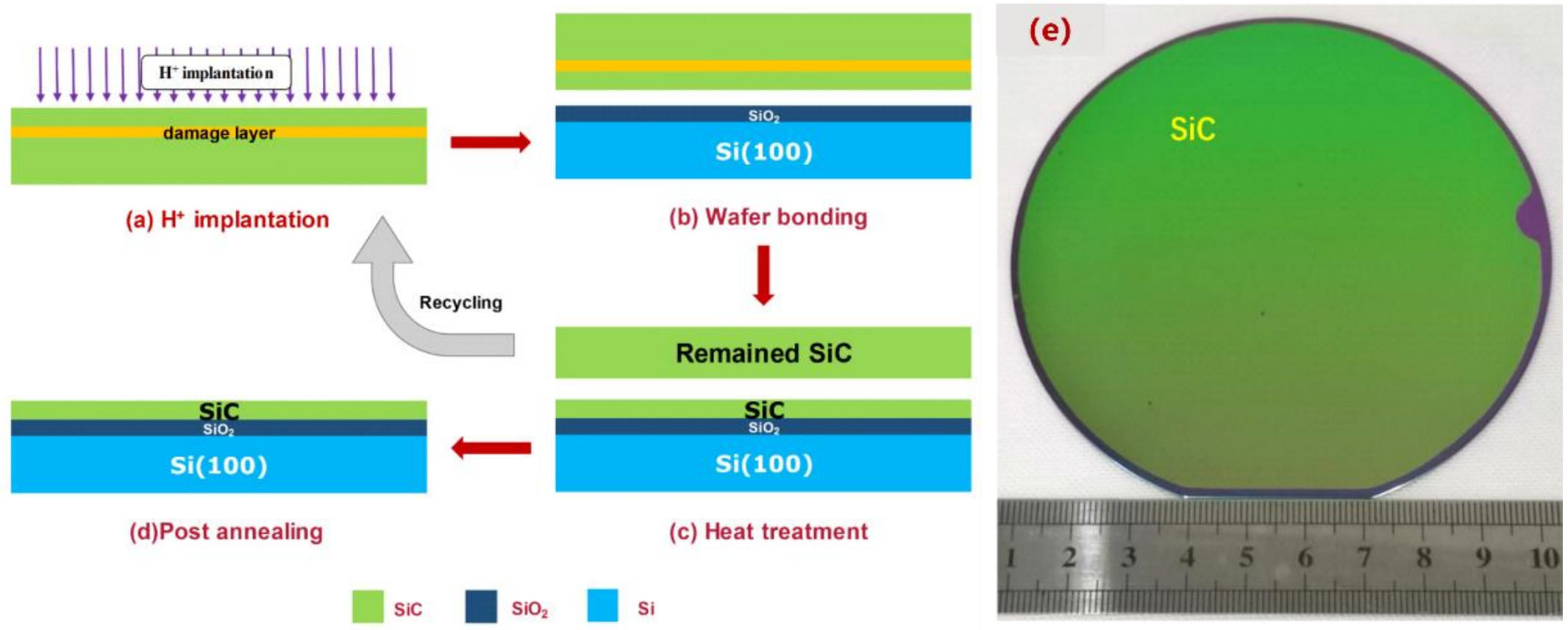

**Figure 3.** Heterogeneous integration of SiC film with Si (100) substrate using the ion-cutting method. (a) Implanting $H^+$ ions into the SiC wafer; (b) Cleaning and bonding SiC with Si (100) handle wafer; (c) Annealing and transferring the SiC film onto the Si (100) substrate; (d) Post-annealing to enhance the bonding strength and to recover the damage and defect due to the implantation. (e) Photograph of 4-inch wafer-scale 4H-

SiCOI substrate with a thickness deviation about +/- 0.2%. Reprinted with the permission from A. Yi *et al.,* Opt. Mater. **107**, 109990 (2020). Copyright 2020 Elsevier[74].

Theoretically, the depth and distribution of $H^+$ could be approximately calculated by "Stopping and Range of Ions in Matter" (SRIM) simulation as shown in Fig.4[70]. The implanted $H^+$ ions generally follow a Gaussian distribution in the 4H-SiC bulk crystal. The depth of the damage layer is adjacent to the position of the Gaussian peak, which can be accurately controlled by the implantation energy[73,76]. During the thermal treatment process, implanted $H^+$ ions continuously assembled into the damage layer and then caused the growth of bubbles. Finally, SiC film can be exfoliated with an accurately uniform thickness due to the precise splitting depth of the expansive bubbles[77]. By using the parameters shown in Fig. 4, the depth of the Gaussian peak was estimated to be about 1.2 μm away from the 4H-SiC top surface at the implantation energy of 180 keV. It should be noted that the simulated results were confirmed in the recent work in which implantation of 170 keV $H^+$ produced a 4H-SiC thin film with a thickness of about 1.1 μm[74].

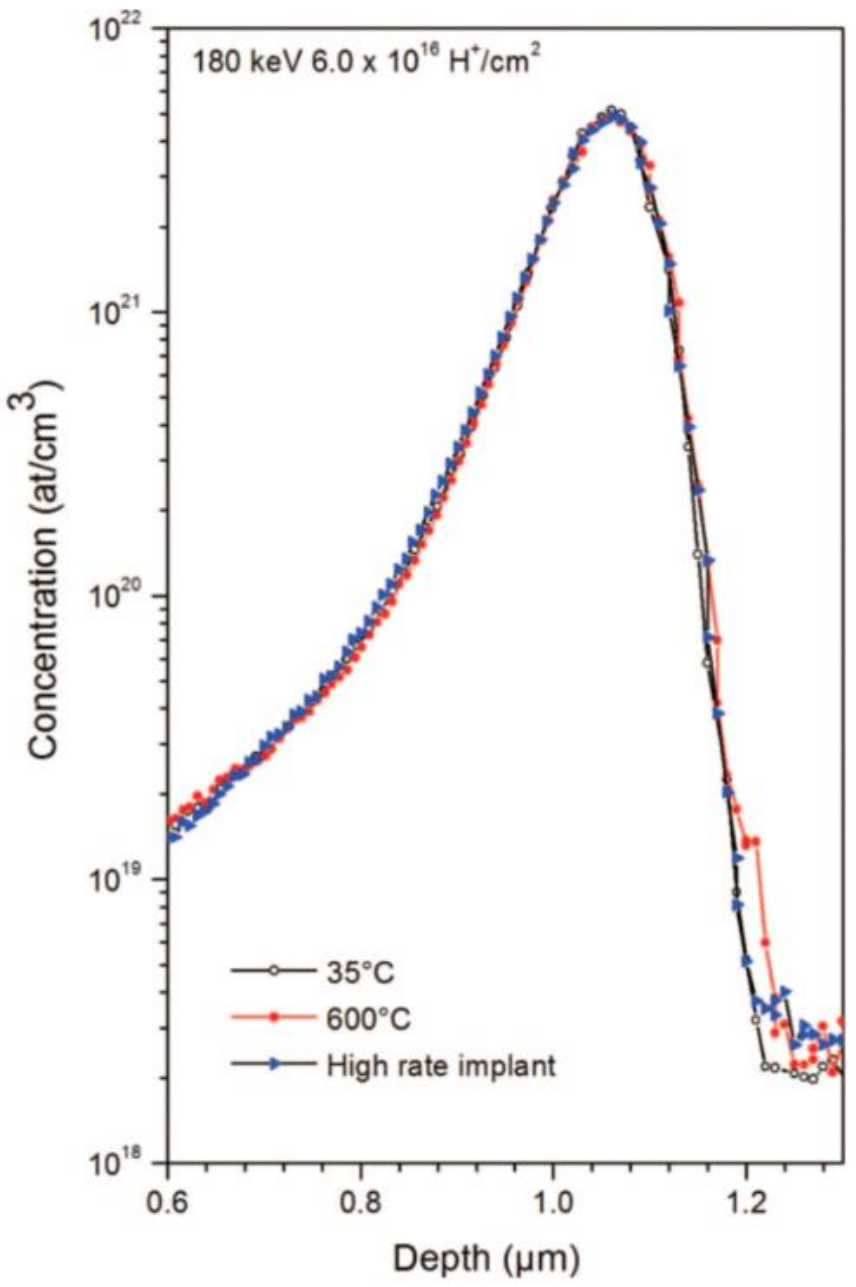

**Figure 4.** SIMS profiles after 180 keV, $6 \times 10^{16}$ $H^+$ $cm^{-2}$ implantation at three temperatures of 4°miscut samples. Reprinted with the permission from V.P. Amarasinghe *et al.,* ECS J. Solid State Sc. **3**, 37 (2014). Copyright The Electrochemical Society[70].

As shown in Fig. 5, the quality of the fabricated 4H-SiCOI film was further characterized by SEM and TEM measurements. A SEM image of the as-prepared 4H-SiCOI substrate after additional annealing and CMP processes was show in Fig. 5(a). The sharp interfaces of $SiC/SiO_2$ and $SiO_2/Si$ could be clearly observed. Figure. 5(b) shows the XTEM image of the $(11\bar{2}0)$ face of the top 4H-SiC thin film. Figure. 5(c)-(f) show the HRTEM images in different regions from top to bottom of the SiC thin film. The amorphous region is about 11 nm from the SiC film surface [see Fig. 5(c)]. This part is mainly composed of Si and O according to the energy dispersive spectrometer (EDS) measurement. The surface of the polished 4H-SiC were prone to oxidization because of the oxidant composition in the slurry during the CMP process. From the HRTEM images in Fig. 5(c)-(f), the periodic lattice fringe was clearly observed from the top, middle, and bottom region of the SiC thin film, suggesting that the SiC thin film persisted a high single-crystalline quality. The single-crystalline quality was confirmed by the selected area electrons diffraction (SAED) shown in the insets of Fig. 5(e)-(f), which illustrated the regular single-crystal diffraction patterns. The poly region about 2 nm among the bonding interface was due to the reaction of Si-OH and the elements diffusion effect between SiC and $SiO_2$ during annealing process. In addition, as the 4H crystal formation, the periodic arrangement of Si and C atoms for four layers circle was observed through the inset STEM images. After removing the top amorphous layer by hydrofluoric acid (HF) treatment, the single-crystalline 4H-SiC thin film with high quality was obtained.

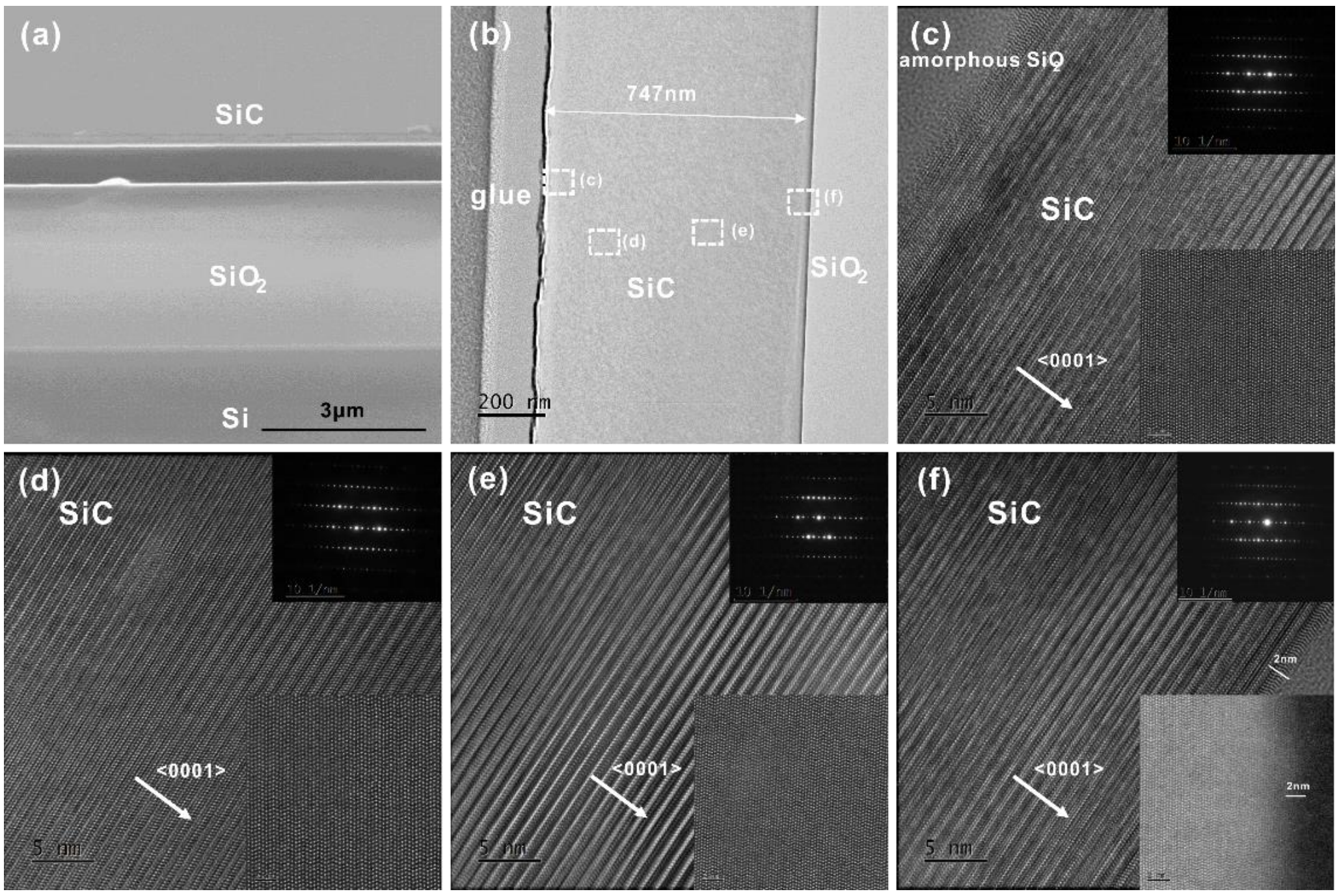

**Figure 5.** The cross sectional characterization of 4H-SiCOI. (a) Cross sectional SEM image of the optimized 4H-SiCOI (post-annealed at 1100 °C after CMP). (b) XTEM image of the optimized 4H-SiCOI. (c)-(f) HRTEM, STEM images and SAED patterns taken from the area marked in (b). Reprinted with the permission from A. Yi *et al.,* Opt. Mater. **107**, 109990 (2020). Copyright 2020 Elsevier[74].

It should be pointed that, the performance of the devices based on ion-cutting 4H-SiCOI is at present limited by the intrinsic absorption loss[37,75]. Unlike the lattice dislocation induced loss in 3C-SiCOI, the point defects caused by $H^+$ implantation are the major factor to the device intrinsic loss which is analyzed by F. Yan *et al.*[40]. Nonetheless, the ion-cutting 4H-SiCOI provides a simpler fabrication process with which extremely high uniformity of the film thickness could be reliably obtained. In general, the ion-cutting 4H-SiCOI platform is suitable for large-scale integrated systems without the requirement of ultra-low loss for the present state. But for the ultra-high *Q* resonators and the single photon source applications, the quality of 4H-SiC film still needs further optimization.

### 2.4.3. 4H-SiCOI prepared by thinning and polishing technique

As an alternative approach, researchers have developed bonding, thinning and polishing techniques to produce the thin film of 4H-SiC on $SiO_2$/Si. In the past few years, S. Ko *et al.*, D. Lukin *et al.* and B.S. Song *et al.* reported their works based on this platform[26,32,45,78]. This process supports 4H-SiC thin film with a high yield and pristine crystal quality, and can be scaled up to wafer size for manufacturing. With this material, nonlinear photonic devices could approach the limitation of an intrinsic property of high-purity 4H-SiCOI. Moreover, the *Q* factors reported in the works of are as high as about $5 \times 10^6$, showing a more brilliant device performance than previous reports up to now[26,30,45].

The fabrication process based on thinning and polishing technique is schematically shown in Fig. 6(a). (0001)-cut 4H-SiC and Si wafers were treated by RCA, following by thermally wet oxidization at 1100 ℃. After an $O_2$/$N_2$ plasma activation on the surface, the two wafers were bonded to each other at room temperature and then annealed at 250 ℃. Subsequently, the bonded structure was annealed again at 1100 ℃ to permanently dehydrate and dehydroxylate the buried oxide (BOX) layer so as to enhance the bonding strength to about 2000 mJ $m^{-2}$. After annealing process, the bonded wafer was further grinded and polished down to 100 μm with a surface roughness of around 1nm. Although the thickness uniformity was somehow reduced, high pristine crystal quality of the 4H-SiC thin film could be attained by using this unique method. Fig. 6(b) shows a 4 inch 4H-SiCOI substrate prepared by the thinning and polishing technique[30]. With a careful treatment, a relatively low thickness deviation of the 4H-SiCOI substrate was obtained. The fraction of uniform area within the thickness range of 2 to 4 μm exceeds 60% was illustrated in Fig. 6(c). Due to the inhomogeneity of the film thickness, the wafer was then cut into $10 \times 12$ $mm^2$ dies, and each die was further thinned down to the desired thickness by inductively coupled-plasma reactive-ion-etching (ICP-RIE) in $SF_6$/$O_2$ plasma and CMP treatment. Figure. 6(d) shows a photograph of a $10 \times 12$ $mm^2$ die with the thickness of $800 \pm 80$ nm. The uniformity of the large area ($10 \times 12$ $mm^2$ die) is sufficient for proof-of-principle demonstrations of high-performance photonic cavities whose quality factors were already approaching the intrinsic limit. For long-term run, further optimization of the thickness uniformity towards industry-standard fabrication of photonic circuits are required.

In general, the incompatibility between the film crystal quality (thinning SiCOI) and the large-scale thickness uniformity (ion-cutting SiCOI) is the major trade-off in 4H-SiC photonics. To overcome the fabrication impediment, 4H-SiCOI would have great potential for the next generation integrated photonics platform after SOI and LNOI due to its excellent intrinsic optics, quantum properties and CMOS compatibility.

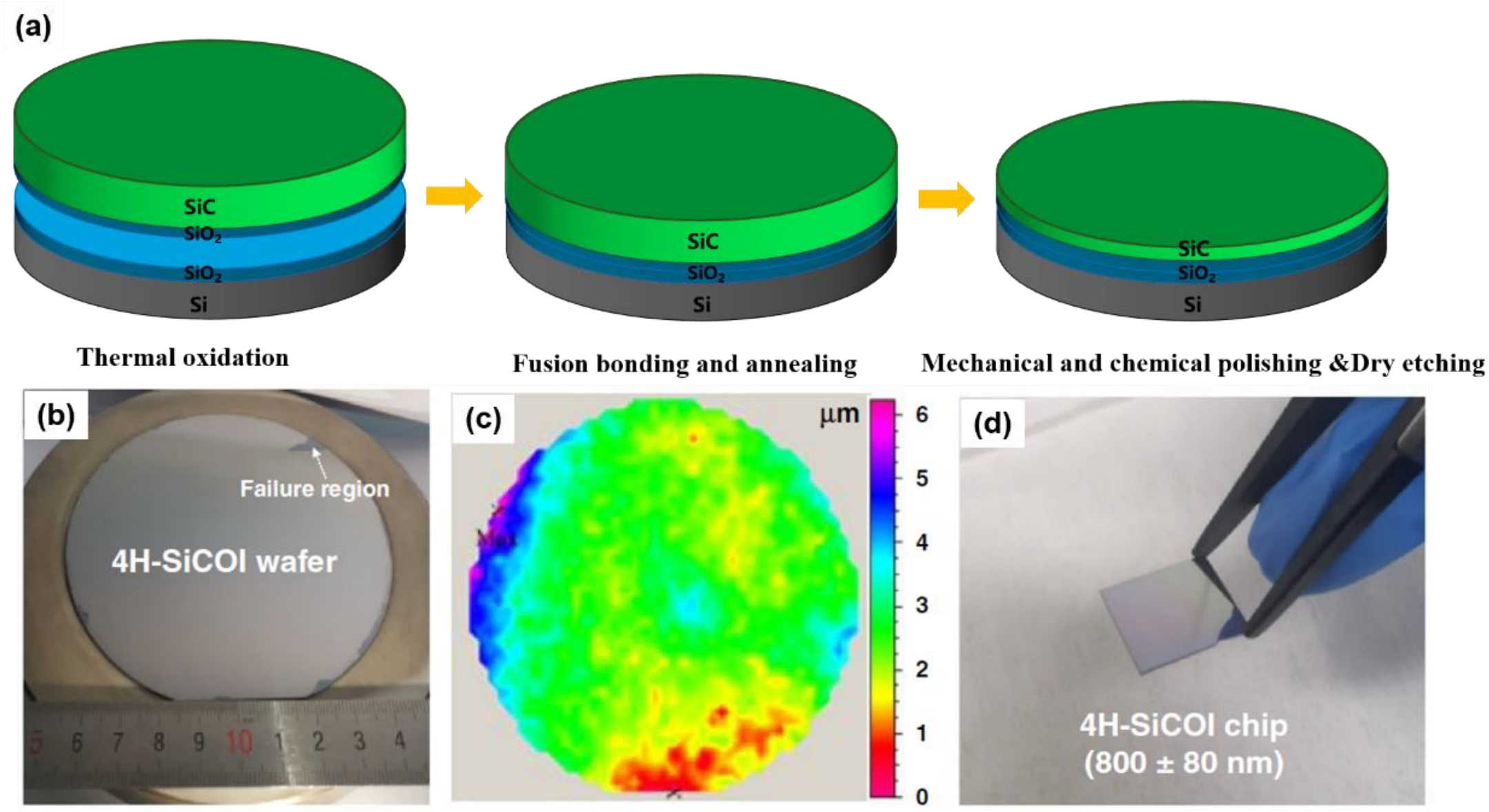

**Figure 6.** Fabrication process flow of single crystalline 4H-SiCOI substrate. (a) Thermal oxidation of SiC and Si wafers. (b) Photograph of a 4-inch wafer-scale 4H-SiCOI substrate fabricated using bonding and thinning method, the failure region is marked. (c) Total thickness variation of the 4H-SiCOI substrate. (d) Image of a 4H-SiCOI die. Reprinted with the permission from C. Wang *et al.,* Light Sci. Appl. **10**, 139 (2021). Copyright C. Wang *et al.* 2021[30].

## 3. Photonic devices on SiC material platform

In order to explore the application of SiC PICs, reliable photonic components need to be developed, including low-loss waveguides that transmit optical signals, and high-$Q$ optical cavities that enhance optical field strength. These elements are the building blocks of PICs. This chapter reviews the development of basic integrated photonic components on SiC platforms, including optical resonators and waveguides. The main properties of the reported optical elements are summarized.

### 3.1 Optical resonant cavity

Optical resonant cavities, including micro-rings, micro-disks, and photonic crystal nano-cavities can confine light inside a micro-volume for long periods of time, which can strongly enhance light-matter interaction, making it an ideal structure for SiC platforms to realize highly efficient nonlinear optical conversion[79], optomechanics interaction[80] and quantum information processing[81]. Enhancement through light-matter interaction in a micro-resonator is quantified by the Purcell factor[82], given by $F = Q/V$, where $Q$ is the quality-factor, and $V$ is the mode volume. $Q$ and $V$ respectively characterize the resonant cavity's ability to confine light in time-field and space-field, the key challenge of the resonant cavity is to fabricate a high-$Q$ and a small mode volume cavity. Fabrication of resonant cavity in SiC is typically achieved by dry-etching. In the case of a fabricated resonant cavity, the $Q$ factor is generally comprised of three contributions: scattering loss, intrinsic materials absorption and radiation loss. Scattering loss is mainly induced by the roughness of side wall of a cavity. A lot of works have been reported to optimize the SiC etching and oxidation process in order to realize smooth waveguide wall[67,83]. Basically, absorption loss originates from an imperfect lattice structure. For a long period of time, because many types of defects are unavoidably introduced into SiC lattice during thin film preparation, the intrinsic absorption loss has been a main loss in SiC optical waveguide[67,84]. SiC thin film has different sources of intrinsic loss corresponding to different preparation processes, such as ion-cutting process, epitaxial growth or deposition method. This section reviews several types of optical resonant cavities and their performance in SiC platforms.

### 3.1.1 Whispering-gallery modes (WGMs) resonators

The simplest form of optical cavities realized in SiC is either a ring or a cylindrical-disk. These structures can support the confinement of light by internal reflections from the circular edge of the resonator, which is called the WGMs. Table 2 summarizes the benchmark works of WGMs-based micro-disk or micro-ring resonators in SiC. In the early demonstrations, thanks to the mature process of growing 3C-SiC on Si, most micro-resonators such as micro-disks[48,53,85,86,87] with $Q$-factors of $2.3 \times 10^3$, $6.19 \times 10^3$, $1.4 \times 10^4$, $3.83 \times 10^4$, $5.12 \times 10^4$, and micro-ring[88] with a $Q$-factor of $1.41 \times 10^4$ were realized on 3C-SiC/Si platforms. Generally, these micro-resonators were patterned by electron-beam lithography tool and etched by RIE-ICP method, followed by undercutting the Si substrate using $XeF_2$ dry etching or KOH wet etching methods. The fabricated micro-disk with a high intrinsic $Q$-factor of $5.12 \times 10^4$ paves the way for strong coupling between single defect points and WGMs[48]. It should be noted that lower $Q$-factors in some works[49,85,89] are due to targeting the visible spectral region and the smaller radius around one-micron scale, which is designed to couple SiC color center in these cavities. When the target band is in the visible spectral region, the resonator is more susceptible to the fabrication imperfection. Otherwise, a smaller radius leads to relatively higher bending loss. The best $Q$-factor reported in 3C-SiC so far for micro-ring resonator is $2.42 \times 10^5$ on the 3C-SiCOI platform[84]. The detailed process flow is described in the previous section. The authors also demonstrated recorded high-$Q$ factors at near-visible and visible wavelengths, of $1.12 \times 10^5$ and $8.3\times 10^4$ of 770 nm and 650 nm, respectively.

**Table 2**. WGMs resonators of various SiC platforms

| Material | preparation technics | resonator type | $Q_{exp}$ | Year of publication |
|---|---|---|---|---|
| **3C-SiC-on-Si** | epitaxially grown on Si | micro-disk | $6.19\times10^3$ (1553.1 nm) | 2013[53] |
| **3C-SiC-on-Si** | epitaxially grown on Si | micro-ring | $1.41\times10^4$ (1543.2 nm) | 2013[88] |
| ***p*-type 4H-SiC** | epitaxially grown on n-4H-SiC | micro-disk | $9.2\times10^3$ (617.5 nm) | 2014[49] |
| ***a*-SiC-on-Si** | PECVD | micro-disk | $1.3\times10^5$ (1545 nm) | 2014[55] |
| **3C-SiC-on-Si** | hetero-epitaxial grown on Si | micro-disk | $5.12\times10^4$ (1551 nm) | 2014[48] |
| **3C-SiC-on-Si** | two-step CVD | micro-disk | $2.3\times10^3$ (769 nm) | 2015[85] |
| **3C-SiC-on-Si** | epitaxially grown on Si | micro-disk | $3.83\times10^4$ (1553 nm) | 2015[86] |
| **3C-SiC-on-Si** | epitaxially grown on Si | micro-disk | $1.9\times10^3$ (640 nm) | 2017[57] |
| **3C-SiC-on-Si** | epitaxially grown on Si | micro-ring | $2.4\times10^4$ (1527.5 nm) | 2017[87] |
| **3C-SiCOI** | epitaxy, bonding, and etching thinning | micro-ring | $1.42\times10^5$ (1561.3 nm) | 2018[43] |
| **3C-SiCOI** | epitaxy, bonding, and etching thinning | micro-ring | $1.39\times10^5$ (1561.3 nm) | 2019[90] |
| ***a*-SiCOI** | PECVD | micro-ring | $1.6\times10^5$ (1568.2 nm) | 2019[91] |
| **4H-SiCOI** | ion-cutting process | micro-ring | $7\times10^4$ (1555 nm) | 2019[67] |
| **4H-SiCOI** | ion-cutting process | micro-ring | $7.3\times10^4$ (1549 nm) | 2019[67] |
| **4H-SiCOI** | bonding and thinning | micro-ring | $7.8\times10^5$ (1523 nm)<br>$2.8\times10^5$ (924 nm) | 2019[32,92] |
| ***a*-SiC** | PECVD | double micro-disk | $1\times10^5$ (1537 nm) | 2019[93] |
| **3C-SiC on Si** | hetero-epitaxial grown on Si | micro-ring | $4.10\times10^4$ (1532.7 nm) | 2020[83] |
| **3C-SiCOI** | epitaxy, bonding, and etching thinning | micro-disk | $2.42\times10^5$ (1550 nm);<br>$1.12\times10^5$ (770 nm);<br>$8.3\times10^4$ (650 nm); | 2020[84] |
| **4H-SiCOI** | bonding and thinning | micro-ring | $1.1\times10^6$(1550 nm) | 2020[26] |
| **3C-SiCOI** | epitaxy, bonding, and etching thinning | slot micro-ring | $1.74\times10^4$ (1310 nm); | 2021[94] |
| **4H-SiCOI** | bonding and thinning | micro-ring | $5.61\times10^6$(1550 nm) | 2021[29] |
| **4H-SiCOI** | bonding and thinning | micro-disk | $7.1\times10^6$(1550 nm) | 2021[30] |

Micro-resonators with large bandwidth are critical for achieving the nonlinear and quantum photonics applications. The $Q$ factor of $1.3 \times 10^5$ for micro-disk[55] and $1.6 \times 10^5$ for micro-ring[91] have been achieved on *a*-SiC. Using plasma-enhanced chemical vapor deposition (PEVCD), *a*-SiC can be directly deposited on any materials[95], and has broad prospects of applications for complementary metal oxide semiconductor (CMOS) compatible integrated photonics[91]. Compared to 3C-SiC and *a*-SiC, 4H-SiC has a higher crystal quality due to the homoepitaxial growth and a wider band gap, which has attracted considerable attention. Micro-disk has been first demonstrated in epitaxial *p*-doped SiC on *n*-4H-SiC[49], fabricated by reactive-ion etching (RIE) in $SF_6/O_2$ plasma using alumina microspheres mask, and then the *p*-layer was undercut by a low-damage selective photoelectrochemical wet etch process. A $Q$-factor of $9.2 \times 10^3$ at 617.4 nm has been demonstrated in this work. Very recently, 4H-SiCOI has been achieved by ion-cutting techniques[67] or bonding and thinning methods[26,32], and micro-resonators with $Q$ factors of $7.8 \times 10^5$, $1.1 \times 10^6$, $5.61 \times 10^6$ [29] and $7.1 \times 10^6$ [30] based on 4H-SiCOI have been demonstrated. Direct wafer bonding and thinning method enable 4H-SiCOI to maintain the same optical and quantum qualities as bulk SiC crystal, which provides an ideal platform for integrated nonlinear and quantum photonics in SiC. Details of these applications will be discussed in the later sections.

**Table 3**. Fabricated PhC cavities of SiC

| Material | Preparation Technics | Cavity Type | $Q_{exp}$ | $V_{cavity}$ $(\lambda/n)^3$ | Year of publication |
|---|---|---|---|---|---|
| **6H-SiCOI** | ion-cutting process | 2D hetero-structure | $4.5\times10^3$(1391 nm) | 0.54 | 2011[51] |
| **6H-SiCOI** | ion-cutting process | 2D L3 | $5\times10^2$(420 nm)<br>$1.2\times10^3$(780 nm)<br>$1.1\times10^3$(1480 nm) | /[(a)] | 2011[96] |
| **6H-SiCOI** | ion-cutting process | 2D hetero-structure | $1\times10^4$(1559 nm) | 1.71 | 2012[97] |
| **3C-SiC-on-Si** | epitaxially grown on Si | 2D L3 | $8\times10^2$(1530 nm) | 0.75 | 2013[54] |
| **3C-SiC-on-Si** | epitaxially grown on Si | 2D H1 | $1.5\times10^3$(1128 nm) | 1(H1) | 2014[47] |
| **6H-SiCOI** | ion-cutting process | 2D hetero-structure | $1\times10^4$(1560 nm) | / | 2014[98] |
| ***a*-SiC** | PECVD | 1D nanobeam | $7.7\times10^4$(1536 nm) | 0.6 | 2015[50] |
| ***p*-4H-SiC** | epitaxially grown on n-4H-SiC | 1D nanobeam | $6.7\times10^3$(700 nm) | 0.5 | 2015[64] |
| ***p*-4H-SiC** | epitaxially grown on n-4H-SiC | 1D nanobeam | $5.3\times10^3$(962 nm) | 0.45 | 2017[62] |
| **bulk 4H-SiC** | oblique plasma etching | 1D nanobeam | $4\times10^4$(1509 nm) | 2.23 | 2018[62] |
| **4H-SiC** | bonding and thinning | 2D hetero-structure | $6.3\times10^5$(1514 nm) | 2.1 | 2019[45] |
| **4H-SiC** | bonding and thinning | 1D nanobeam | $1.9\times10^4$(860 nm) | 0.46 | 2019[32] |
| ***p*-4H-SiC** | epitaxially grown on n-4H-SiC | 1D nanobeam | $5\times10^3$(1079 nm) | / | 2020[44] |
| **3C-SiC-on- Si** | PECVD | Zipper | $2.8\times10^3$(1496 nm) | 0.084 | 2020[99] |

[a)] '/' indicates that this value is not shown in the literature.

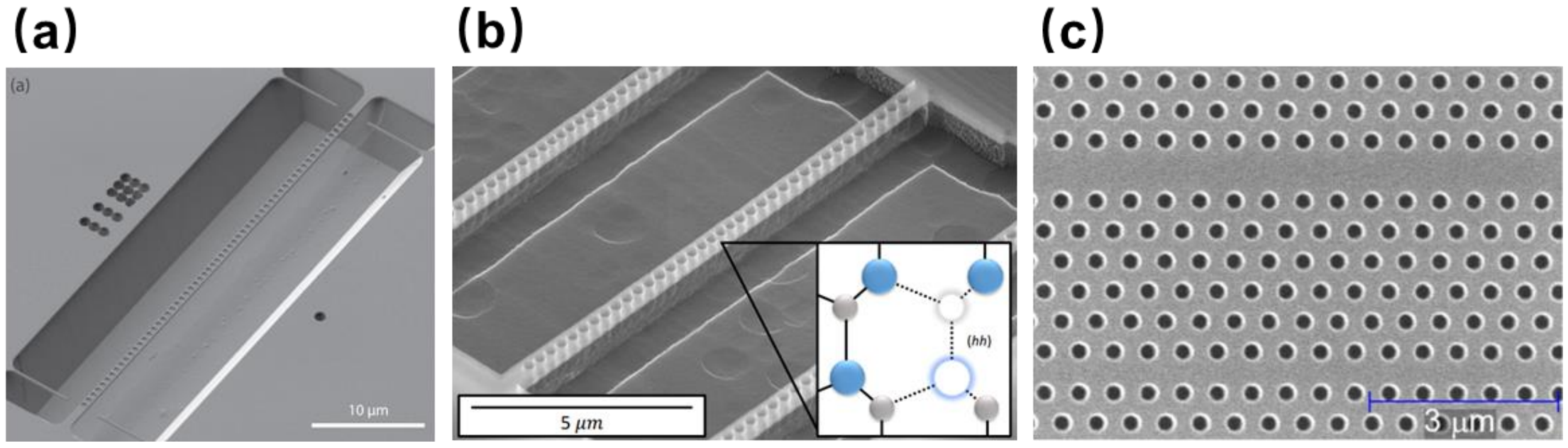

**Figure 7**. SEM images of PhC nano-cavities on various SiC platforms. (a) $a$-SiC photonic-crystal nanobeam cavity. Reprinted with the permission from J.Y. Lee *et al.,* Appl. Phys. Lett. **106**, 041106 (2015). Copyright 2015 AIP Publishing LLC[50]. (b) 4H-SiC nanobeam cavities via dopant-selectively etching method. Reprinted with the permission from A.L. Crook *et al.,* Nano Lett. **20**, 3427 (2020). Copyright 2020 American Chemical Society[44]. (c) 2D PhC nano-cavities on thinning 4H-SiCOI. Reprinted with the permission from B.S. Song *et al.,* Optica **6**, 991 (2019). Copyright 2019 Optical Society of America[45].

### 3.1.2 Photonic crystal nano-cavities

Photonic crystal (PhC) nano-cavities are critical components for PICs[100,101]. Although WGMs-based micro-resonators (such as micro-disk and micro-ring resonators) generally have the highest $Q$, the mode volume is also large, about two orders of magnitude larger than photonic crystal nano-cavities. Thus, to maximize Purcell enhancement in SiC-based resonant cavity, different types of PhC cavities have been realized in different polytypes of SiC. In Table 3, we summarize the previously reported properties of PhC cavities on SiC platforms. In 2011, for the first time ever, a 2D L3 photonic crystals cavity was fabricated based on a 6H-SiCOI wafer[51]. The L3 cavity, composed of three missing holes in a 2D SiC photonic crystal slab structure with a triangular lattice of air holes, was measured a $Q$ factor of $4.5 \times 10^3$ and calculated to a mode volume of $0.54(\lambda/n)^3$, which implies that the optical properties of SiC-based photonic crystal are comparable to those of traditional photonic crystal. Again, the $Q$ factor of $10^4$ was achieved by the same group[97], the authors also investigated the robustness of the ultra-small mode volume SiC-based photonics crystal, and they experimentally demonstrated that SiC-based photonics crystal could significantly suppress multiple photon absorption even at a high input power. 2D L3 cavities based on grown epitaxial 3C-SiC have been demonstrated by M. Radulaski *et al.*[54].The fabrication processes used to obtain photonic crystal nano-cavities on epitaxial 3C-SiC are similar to the waveguide or micro-ring fabrication processes we aforementioned. High-$Q$ 1D-PhC nanobeam cavities with intrinsic optical $Q$ of $7.69 \times 10^4$ and mode volume of $0.6(\lambda/n)^3$ were surprisingly achieved in the $a$-SiC platform[50], as shown in Fig. 7(a). Moreover, B.S. Song *et al.*[62] developed an oblique plasma etching method to fabricate PhC nanobeam in bulk 4H-SiC. Thanks to the optical platform with bulk SiC crystalline integrity, the devices were achieved a high-$Q$ of $4 \times 10^4$. Suspended PhC cavities can be also derived from hexagonal SiC by dopant-selectively photoelectrochemical etching, with $Q$ factors exceeding $5 \times 10^3$ as shown in Fig.

7(b)[44,64,65]. The coupling of single color center to cavity mode has also been demonstrated[44]. Recently, 4H-SiCOI platforms prepared by bulk wafer bonding and thinning technique have garnered much attention. In 2019, B.S. Song *et al.*[45] developed the 4H-SiCOI platform by this technique, achieving a new record $Q$ factor of 6.3 × $10^5$ with the simulated mode volume $2.1(\lambda/n)^3$ in fabricated 2D heterostructure PhC as shown in Fig. 7(c). D. M. Lukin *et al.*[32] demonstrated 1D PhC nanobeam cavities based on 4H-SiCOI platform using similar techniques as B.S. Song *et al.*. The fabricated nanobeam cavities were measured to a $Q$ factor of 1.93 × $10^4$ and calculated a mode vulome of $0.46(\lambda/n)^3$. PhC cavity with such a high-$Q$ factor as well as ultra-small mode volume, will accelerate the research progress of SiC in integrated photonics.

### 3.2 Waveguides

Optical waveguides are the most fundamental components in photonic devices that perform guiding, connecting, switching tasks on a chip. They follow a similar fabrication procedure similar to micro-resonators and their propagation loss can be estimated by the $Q$ factor of a micro-ring resonator. Generally, SiC waveguides are fabricated on the SiCOI platform, in which a SiC thin layer is isolated on a lower-index substrate. In the early research, epitaxial 3C-SiC is the most widely used optical platform thanks to its mature material preparation process[50,53-56]. However, waveguides fabricated on epitaxial 3C-SiC thin film generally have a relatively high intrinsic loss due to the large lattice mismatch between SiC and Si[43]. Previous works in epitaxially 3C-SiC have realized ring resonators with a high intrinsic $Q$ factor up to 2.4 × $10^4$ [88], which indicates the propagation loss around 21 dB $cm^{-1}$. Recently, K. Powell *et al.*[83] achieved loss of 7 dB $cm^{-1}$ by improving the devices fabrication process through high temperature annealing in dry oxygen atmosphere. As a result of annealing the waveguide, a reduction in sidewall roughness from 2.4 nm to below 1.7 nm and an increasing in SiC crystal purity can be achieved, the authors demonstrated a significant reduction in loss from 24 dB $cm^{-1}$ to 7 dB $cm^{-1}$. Nevertheless, these values are close to the limit of intrinsic loss of the epi-3C-SiC. In addition, all photonic components prepared on an epi-3C-SiC/Si platform must be suspended in air due to a higher refractive index Si layer, which suffers from the complex fabrication processes and low reliability.

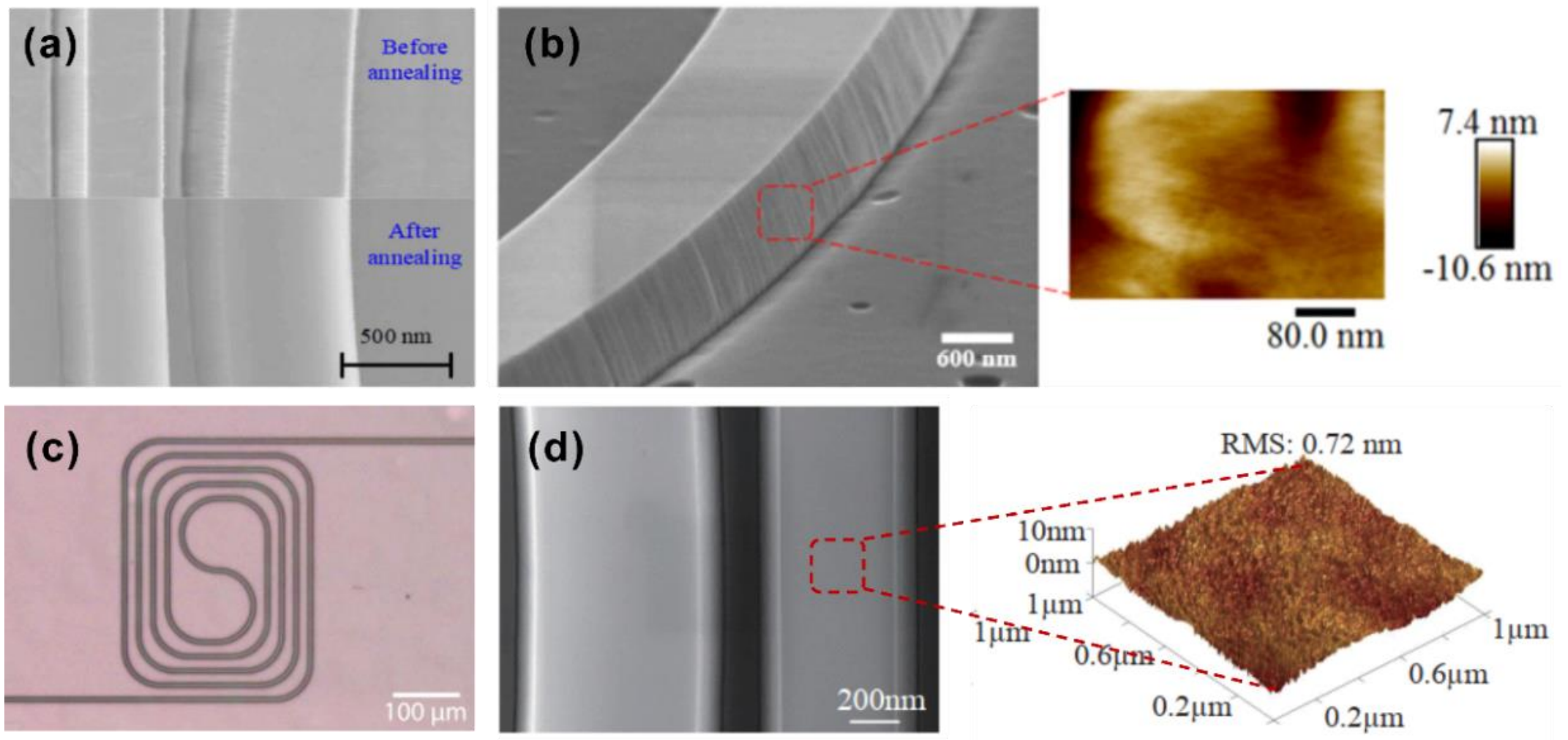

**Figure 8**. Illustrations of several typical reported waveguides. (a) SEM of the waveguide fabricated on epitaxially grown 3C-SiC on Si by K. Powell *et al.*. Reprinted with the permission from K. Powell *et al.,* Opt. Express **28**, 4938 (2020). Copyright 2020 Optical Society of America[83]. (b) SEM image of the micro-ring resonator fabricated on 3C-SiCOI platform with 2D AFM image of its sidewall. Reprinted with the permission from T. Fan *et al.,* Opt. Express **26**, 25814 (2018). Copyright 2018 Optical Society of America[43]. (c) Micrograph of SiC waveguide fabricated on 4H-SiCOI platform. Reprinted with the permission from J. Cardenas *et al.,* Opt. Lett. **40**, 4138 (2015). Copyright 2015 Optical Society of America[75]. (d) SEM image of the waveguide fabricated on 4H-SiCOI platform with 2D AFM image of its top surface. Reprinted with the permission from Y. Zheng *et al.,* Opt. Express **27**, 13053 (2019). Copyright 2019 Optical Society of America[67].

Adibi's group recently transferred epitaxially grown 3C-SiC thin film onto thick silicon oxide isolation layer through the bonding process and chemical-mechanical polishing technique[43,84,90]. The detailed process flow is described in section 2.4.1. The propagation loss extracted from a high-$Q$ micro-ring ($1.42 \times 10^5$) on the 3C-SiCOI platform is around 2.9 dB $cm^{-1}$ shown in Fig. 8(b). Compared with previous works in epi-3C-SiC/Si or deposited *a*-SiC, significant improvement in reducing losses have been achieved. Later in 2015, waveguide fabricated in crystalline 4H-SiCOI was reported and the propagation loss was measured to be about 7 dB $cm^{-1}$. Recently, thanks to the improvement of SiC etching, polishing and wet-oxidation processes, a high-$Q$ factor of $7.3 \times 10^4$ micro-ring resonators has been obtained[67]. Very recently, the high quality crystalline 4H-SiCOI platform has been developed in wafer-scale[74], which could promote the further application of 4H-SiCOI in large-scale integrated photonics devices. However, it is believed that the propagation loss of current waveguide devices based on SiCOI (3C-SiCOI or 4H-SiCOI) prepared by ion-cutting technique is mainly attributed to the material intrinsic absorption[67].

Another solution to produce pristine SiC thin film has been developed. The detailed process flow is described in section 2.4.3. A recorded high-$Q$ of $7.8 \times 10^5$ micro-ring resonator was demonstrated, corresponding to an ultra-low propagation loss of 0.5 dB $cm^{-1}$ [32]. What's more[26], the waveguide loss of 0.38 dB $cm^{-1}$ has been demonstrated, and this is close to the intrinsic material loss of the high-purity bulk 4H-SiC wafers (0.3 dB $cm^{-1}$)[75]. In their works, the intrinsic absorption of a sublimation-grown bulk 4H-SiC was measured to be 0.02 dB $cm^{-1}$. This value indicates that it is possible to achieve lower loss in SiC waveguide. The ultra-low loss SiC waveguide will stimulate a renewed interest in SiC photonic devices, for example, nonlinear optics, optomechanics, and quantum photonics.

## 4. Nonlinear optical elements

Nonlinear optics shows great potential for a number of applications such as high-speed signal processing[102], frequency conversion[17,103], sensing[104]. SiC offers both second and third order optical nonlinearities. Besides, with the attractive optically active spin qubits, SiC has emerged as a contender not only for conventional nonlinear photonics, but also for spin-based quantum photonic applications[26,32]. In this chapter, we review the recent advances of SiC-based nonlinearities, including dispersion engineering, second-harmonic generation (SHG), electro-optics, four-wave mixing (FWM) and optical parametric generation.

## 4.1 Dispersion engineering

Engineering of group velocity (GV) and group velocity dispersion (GVD) for the micro-resonator or waveguide is an indispensable task in nonlinear optics, and the quantum application of point defects does not necessarily need dispersion engineering. The group velocity dispersion (GVD) can be expressed by

$$\beta_2 = \frac{1}{c}\frac{\partial n_g}{\partial \omega},$$

where $c$ is the speed of light, $\omega$ and $n_g$ are the angular frequency and the group velocity of optical wave respectively. Notably, the dispersion can be classified into types: material and geometric dispersion[105,106]. The geometric dispersion can be further divided into the dispersion related to the waveguide cross-section and the dispersion associated with the micro-resonator curvature. Similar to the most materials, SiC exhibits normal dispersion ($\beta_2 < 0$) at the telecom wavelengths, and hence the dimensions of the waveguide or resonators should be carefully designed in order to compensate the material dispersion in a proper way.

The dispersion of a mode family with resonance frequency $\omega_\mu$ can be approximated by using the Taylor expression[107]

$$\omega_\mu = \omega_0 + \mu D_1 + \frac{\mu^2}{2} D_2 + \frac{\mu^3}{6} D_3 + \cdots,$$

where $\mu$ dominates the relative mode number, $D_1$represents the free-spectral range around $\omega_0$[108] and $D_2$ is related to the GVD parameter $\beta_2$ as $D_2 = -\frac{c}{n_g} D_1^2 \beta_2$, which is the derivative of group delay per unit transmission length with respect to the wavelength. If $D_2 > 0$ ($D_2 < 0$), the dispersion is said to be “anomalous” (“normal”). It has long been known that optical parametric oscillators (OPOs) and broadband Kerr comb generation (soliton formation) occur only if the group velocity dispersion is anomalous[108,109] (additional possibilities are available if the wave is allowed to interact with a second, distinguishable wave through cross-phase modulation), because the broadening caused by the self-phase modulation (SPM) effect experienced by the pulse requires a proper anomalous dispersion effect to counteract.

It is showed that dispersion engineering could be well achieved by adjusting the cross-sectional profiles of the SiC waveguide[67]. By tuning the width of the waveguides, the GVD can be tailored from the normal dispersion regime to the anomalous dispersion regime. These observations indicate that there are multiple high-$Q$ mode families with anomalous dispersions. They can potentially be employed for optical parametric oscillation and broadband Kerr comb generation.

### 4.2 Second-harmonic generation in SiCOI

The three most commonly used polytypes, 3C-SiC ($\bar{4}3m$), 4H-SiC ($6mm$) and 6H-SiC ($6mm$) exhibit non-zero second order susceptibilities $\chi^{(2)}$ due to the lack of centrosymmetric lattice structure. Although the second-order nonlinearity of SiC was predicted and measured in 1892[110,111], it was not clearly investigated until 2014 that the SHG effect was demonstrated by using 6H-SiC PhC cavity[98]. In this demonstration, L3 PhC cavity with a $Q$-factor of $1 \times 10^4$ for the fundamental mode was prepared. The conversion efficiency of the SHG in the cavity was estimated to be 15% $W^{-1}$ at a maximal input power of 0.17 mW. Instead of using bulk material, the same group has enabled an unprecedented increase of the SHG conversion efficiency (1900% $W^{-1}$) using SiCOI in 2019[45]. This achievement represents a remarkable progress in SiC nonlinear photonics.

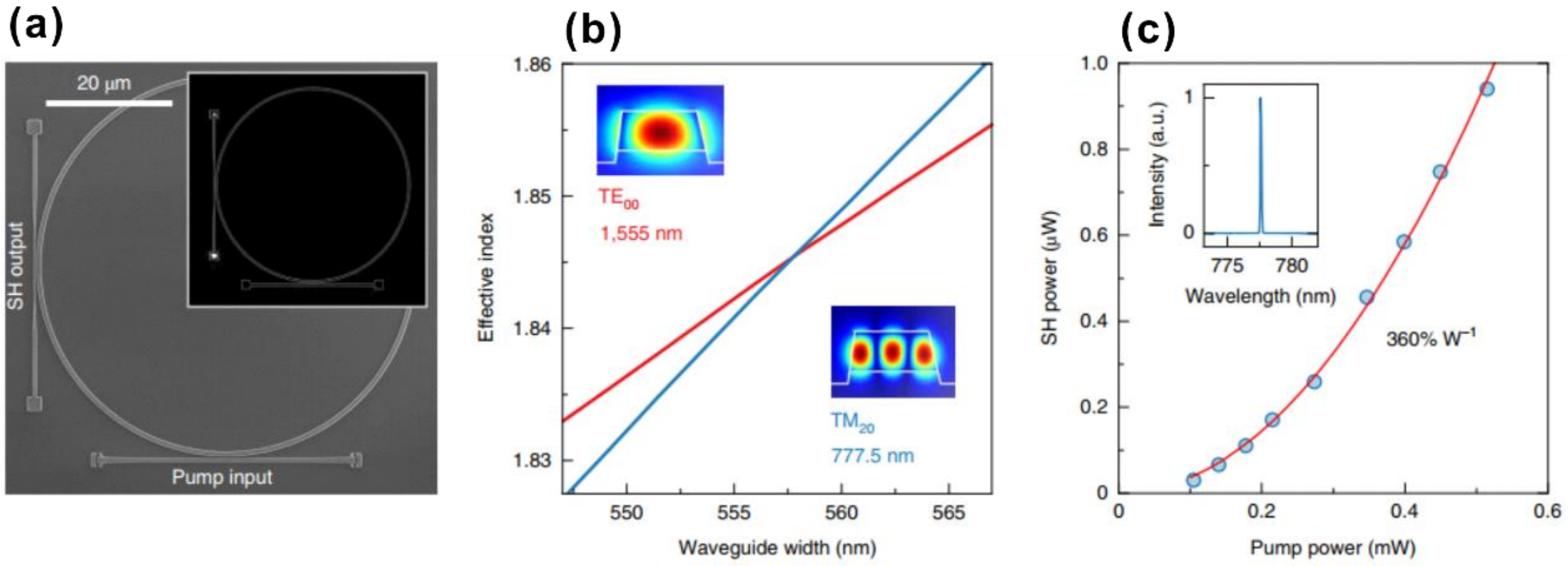

**Figure 9**. Efficient second-order frequency conversion in a micro-ring resonator on 4H-SiCOI platform prepared by the thinning and polishing technique. (a) SEM image of a fabricated ring resonator. Inset: optical image of the ring resonator. (b) Calculation of the phase-matching condition for the 1,555 nm $TE_{00}$ and 777.5 nm $TM_{20}$ modes, indicating that phase-matching can be satisfied when the waveguide width is 560 nm. Insets: Calculated electric field intensity distribution for the $TE_{00}$ and $TM_{20}$ modes. (c) Dependence of SHG power on the fundamental power. The red fitting line reveals a normalized conversion efficiency of 360% $W^{-1}$. Inset: the second-harmonic signal imaged on a spectrometer. Reprinted with the permission from D.M. Lukin *et al.,* Nat. Photonics **14**, 330 (2019). Copyright 2019 Springer Nature Limited[32].

Unlike PhC cavities, efficient nonlinear frequency conversion in waveguides or WGMs-based optical resonators must satisfy the critical phase matching condition[112]. For most single-crystal materials, the dispersion relationship between energy and momentum is not linear due to the intrinsic material dispersion, so the phase matching conditions are very difficult to be satisfied. However, as an alternative approach, quasi-phase matching can be realized by dispersion engineering[113-115]. D.M. Lukin *et al.*[32] achieved efficient frequency conversion with SHG conversion efficiency of 360% $W^{-1}$ in an ultra-high-$Q$ micro-ring resonator. By adjusting the cross-sectional dimensions of the waveguide, a fundamental quasi-TE mode at 1555 nm and a quasi-transverse-magnetic mode at 777.5 nm can have the same effective refractive index in a SiC nano-photonic waveguide, fulfilling the requirements of phase matching condition, as shown in Fig. 9(b). High brightness quantum emitters were also demonstrated in this work, in a sense, the 4H-SiCOI platform was proved that strong spin qubit emitters and efficient nonlinear optics could be simultaneously achieved by a monolithic approach. This makes 4H-SiCOI a major candidate for spin-based quantum photonics, with simultaneous monolithic integration and efficient quantum frequency conversion.

### 4.3 Electro-optic modulators

The intrinsic electro-optic (EO) effect is a phenomenon which the variation of material refractive index depends linearly on the strength of the applied electric field, also known as the Pockels effect. This effect enables realizing resonance frequency shifting, electro-optic (EO) modulators and electro-optic comb generation. Particularly, the EO modulator, enabling the translation between the electronic signal and the optical signal, is a key component for the modern telecommunication system. Electric-optic modulators have ever been demonstrated on a variety of platforms, such as silicon, indium phosphide, graphene and polymers[116]. However, these modulators have shown many drawbacks and could not completely address the need of electro-optic systems, especially for the next generation of photonic applications. For example, Si modulators rely on the free carrier dispersion effect, but this effect is intrinsically absorptive and nonlinear, leading to a degradation and signal distortion for the optical modulation amplitude. Alternatively, non-centrosymmetric materials with the EO effect (Pockels effect) provide a viable solution, allowing the refractive index to be changed linearly with respect to the applied electric field. In recent years, LN has become a promising platform for high-performance EO modulators featuring with a CMOS-compatible voltage[13] due to its large electro-optic coefficient ($r_{33}$~27 pm $V^{-1}$). However, LN is incompatible with CMOS fabrication processes. The fabrication of nano-photonic structures has been proven to be difficult, which renders LN inconvenient in practical photonic applications. In contrast, SiC is CMOS-compatible and it simultaneously possesses large EO coefficient so that fast EO modulators fabricated on SiC material platforms can be anticipated[38,117].

The linear EO effect can be usually described by a two-dimensional matrix $r_{hk}$, known as EO coefficients. For 3C-SiC with a point group of $\bar{4}3m$, the EO coefficients can be reduced to a simple form.

$$r_{hk} = \begin{pmatrix} 0 & 0 & 0 \\ 0 & 0 & 0 \\ 0 & 0 & 0 \\ r_{41} & 0 & 0 \\ 0 & r_{52} & 0 \\ 0 & 0 & r_{63} \end{pmatrix},$$

where, $r_{41}$, $r_{52}$, $r_{63}$ have the equal value.

For 4H-SiC or 6H-SiC with a point group of $6mm$, the form of EO coefficients is different,

$$
r_{hk} = \begin{pmatrix} 0 & 0 & r_{13} \\ 0 & 0 & r_{23} \\ 0 & 0 & r_{33} \\ 0 & r_{42} & 0 \\ r_{51} & 0 & 0 \\ 0 & 0 & 0 \end{pmatrix},
$$

where $r_{13} = r_{23}$, $r_{51} = r_{42}$.

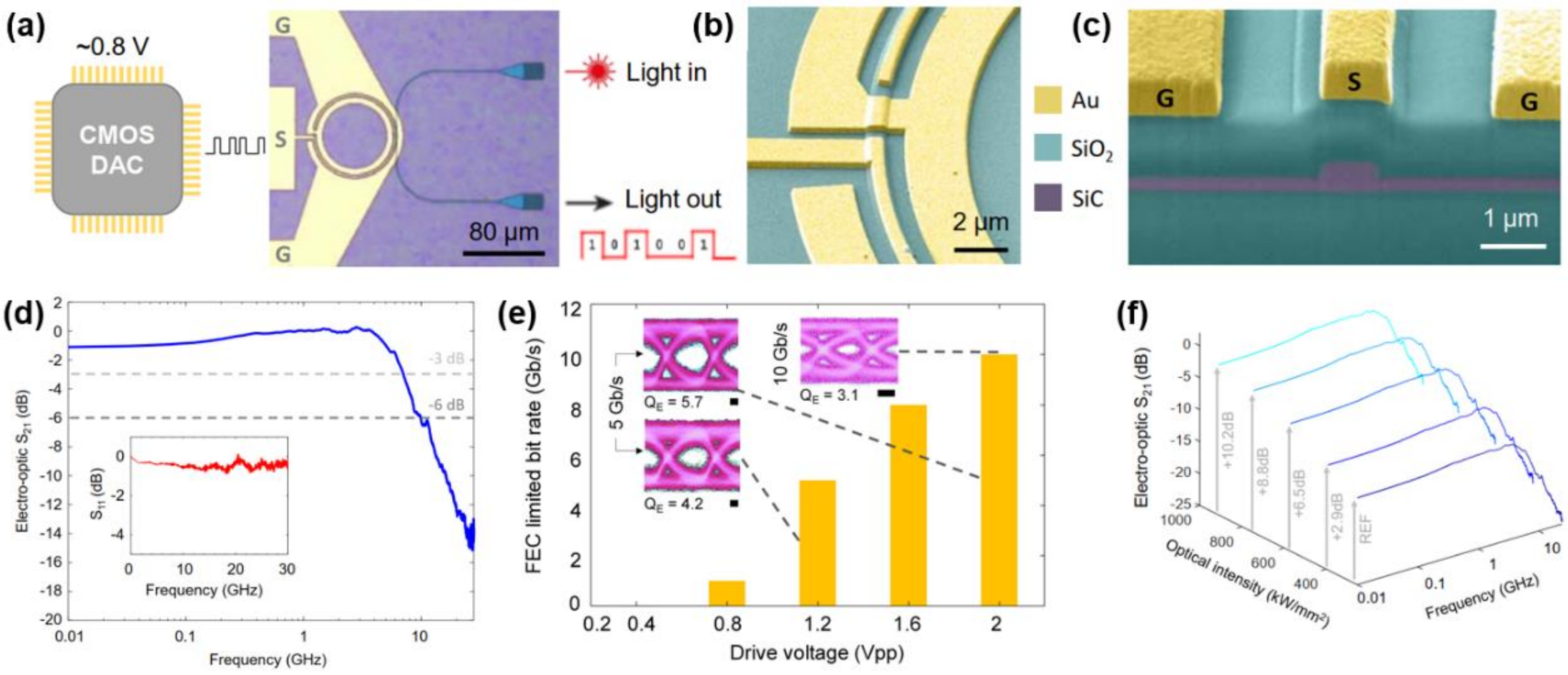

**Figure 10**. Integrated Pockels modulator in 3C-SiCOI. (a) Overview of the fabricated 3C-SiC modulator showing compatibility with CMOS voltages. (b) SEM image of the modulator electrodes and (c) Its cross-section structure. (d) Modulator bandwidth characterization. RF s-parameter characterization featuring a -3dB and -6dB bandwidths of 7.1 GHz and 9.9 GHz respectively. (e) Eye diagram *Q* factors with increasing the drive voltage. (f) High power operation characterization. The RF responses shows an improvement with increasing the optical intensities. Reprinted with the permission from K. Powell *et al.,* Nat. Commun. **13**, 1851 (2022). Copyright K. Powell *et al.* 2022[117].

Despite the potential for EO applications, there have been few experimental demonstrations on SiC EO modulators. In 1991, a LPCVD 3C-SiC thin film on Si substrate was used to investigate the EO effect with an EO coefficient $r_{41}$ (~2.7 pm V$^{-1}$) at 633 nm. This first experimental achievement opens the door for high optical density in SiC modulators. Later in 2000, some schemes of 3C-SiC-based Pockels modulators were proposed. Recently, a racetrack micro-resonator EO phase-shifter on 3C-SiC platform was demonstrated in which a voltage-length product ($V_\pi \cdot L$) of 118 V · cm, corresponding to an EO coefficient of ~2.6 pm V$^{-1}$ [38]. At the same time, a CMOS-compatible micro-ring EO modulator on 3C-SiC was also demonstrated and an EO coefficient of ~1.5 pm V$^{-1}$ was extracted[117]. As shown in Fig. 10(a) - (c), electrodes of the SiC modulator consisted of a pair of ground electrodes placed next to the sides of the waveguide and a signal electrode above the waveguide. The modulator supported a maximum bit rate up to 10 Gb/s and an extinction ratio of 3 dB, as shown in Fig. 10(d). The low voltage operation was demonstrated in Fig. 10(e), and the drive voltage could be reduced from $V_{pp}$=2 V to 1.2 V while maintaining 140 an open 5 Gb s$^{-1}$ non-return-to-zero eye diagram. Moreover, as shown in Fig. 10(f), the CMOS-compatible modulator featured a low-noise modulation at high optical intensities up to 913 kW $mm^{-2}$, indicating a key advantage of a SiC-based EO modulator. The demonstration of CMOS-compatible EO modulator is a significant milestone in the development of SiC PICs, paving the way for monolithic integration of modulators, nonlinear elements and spin defect.

As the development of SiC photonics continues, 4H-SiC is gradually attracting extensive research attention. When an electric field is applied along the *c*-axis on an (1000)-oriented 4H-SiC thin film, it changes the index in *a*-, *b*- and *c*-axis but causes no rotation of principal axes, which simplifies the choices of the orientation of electro-optic devices and waveguide polarization mode as compared to 3C-SiC. The EO application on 4H-SiC platform remains unexplored as yet. Thus far, the performance of the integrated 4H-SiC electro-optic modulator is still unclear.

### 4.4 Third-order nonlinearity

Compared with the second-order nonlinear process, the third-order nonlinear process is rich and complex in a nonlinear media. Dislike the second-order nonlinearity which occurs in materials with central asymmetric structure, the third-order nonlinearity has no dependence on the crystal symmetry. Therefore, the third-order nonlinear processes such as SPM, third-harmonic generation (THG), FWM, OPO and Kerr frequency comb generation have been extensively investigated on different platforms including $Si_3N_4$[118], $LiNbO_3$[119-121], Diamond[17], Silicon[122,123], etc. Among these materials, the unique properties of SiC highlight that it is a promising material for Kerr integrated photonics.

In 2014[55], a strong Kerr nonlinearity phenomenon was characterized on a high-$Q$ *a*-SiC micro-disk resonator. In order to measure the $n_2$ value, a pump-probe self-/cross-phase modulation scheme was developed using two high-$Q$ resonant cavity modes. With an optical $Q$-factor of $1.3 \times 10^5$ at 1550 nm telecom band, its Kerr nonlinearity was characterized through pumping sinusoidally modulated mode at 1545 nm into the resonator. The refractive index was modulated by the optical Kerr effect, which was measured by the probe beam and then the signal was transferred out of the resonator. In this work, a Kerr nonlinear coefficient $n_2$ of $(5.9 \pm 0.7) \times 10^{-15}$ $cm^2$ $W^{-1}$ (1545 nm) was obtained, which is larger than other materials used for Kerr frequency comb generation.

The Kerr nonlinear coefficient can also be measured by carrying out SPM experiments[67,75,91,124]. SPM is the simplest and most important kind of phenomenon in nonlinear optics. That is, due to the optical Kerr effect, when an ultrashort pulse passes through a nonlinear material, it will cause the refractive index of the material to change with the intensity of the different parts of pulse in time domain, resulting in different phases accumulation in different parts of the pulse. As a result, the spectrum corresponding to the pulse will be broadened accordingly for general cases. In SPM experiments, a SiC waveguide was pumped by a laser pulse. After propagating along in the waveguide (generally under normal dispersion), the laser pulse was recorded by optical spectrum analyzer. Subsequently, the Kerr nonlinear coefficient could be extracted using the conversion efficiency equation described by F. Absil *et al.*[125] or using the nonlinear Schrödinger equation (NLSE). With this method, J. Cardenas *et al.*[75] demonstrated strong nonlinearities of $n_2 = 8.6 \pm 1.1 \times 10^{-15}$ cm$^2$ W$^{-1}$ (2360 nm) in 4H-SiC waveguide. The value of $n_2$ is almost three times that of $Si_3N_4$ and on the same order of magnitude as silicon at the same wavelength. Recently, the value of $n_2$ of *a*-SiC around 1550 nm was calculated to be $4.8 \times 10^{-14}$ cm$^2$ W$^{-1}$ [91], about 10 times larger than the earlier result[55]. It should be noted that, in this work, three-photon absorption resulting from the nonlinear loss was observed in *a*-SiC for the first time and the absorption coefficient was characterized to be ~0.01 cm$^3$ GW$^{-2}$.

Four-wave mixing (FWM) has been demonstrated in 3C-SiC and 4H-SiC micro-ring resonators. F. Martini *et al.*[56] achieved a conversion efficiency of 72 dB at a low pump power of 2.9 mW and the nonlinear refractive index $n_2$ was calculated to be $5.31 \pm 0.04 \times 10^{-19}$ m$^2$ W$^{-1}$. Y. Zheng *et al.*[67] reported a higher efficient FWM process in high-$Q$ factor, high-confinement 4H-SiC micro-ring resonators, a FWM conversion efficiency of 21.7 dB was achieved with 79 mW pump power and the nonlinear refractive index ($n_2$) of 4H-SiC was estimated to be $6.0 \pm 0.6 \times 10^{-19}$ m$^2$ W$^{-1}$. The authors also achieved 3 dB FWM conversion bandwidth of more than 130 nm by dispersion engineering. The obtained large FWM bandwidth will be beneficial for a broadband frequency conversion application.

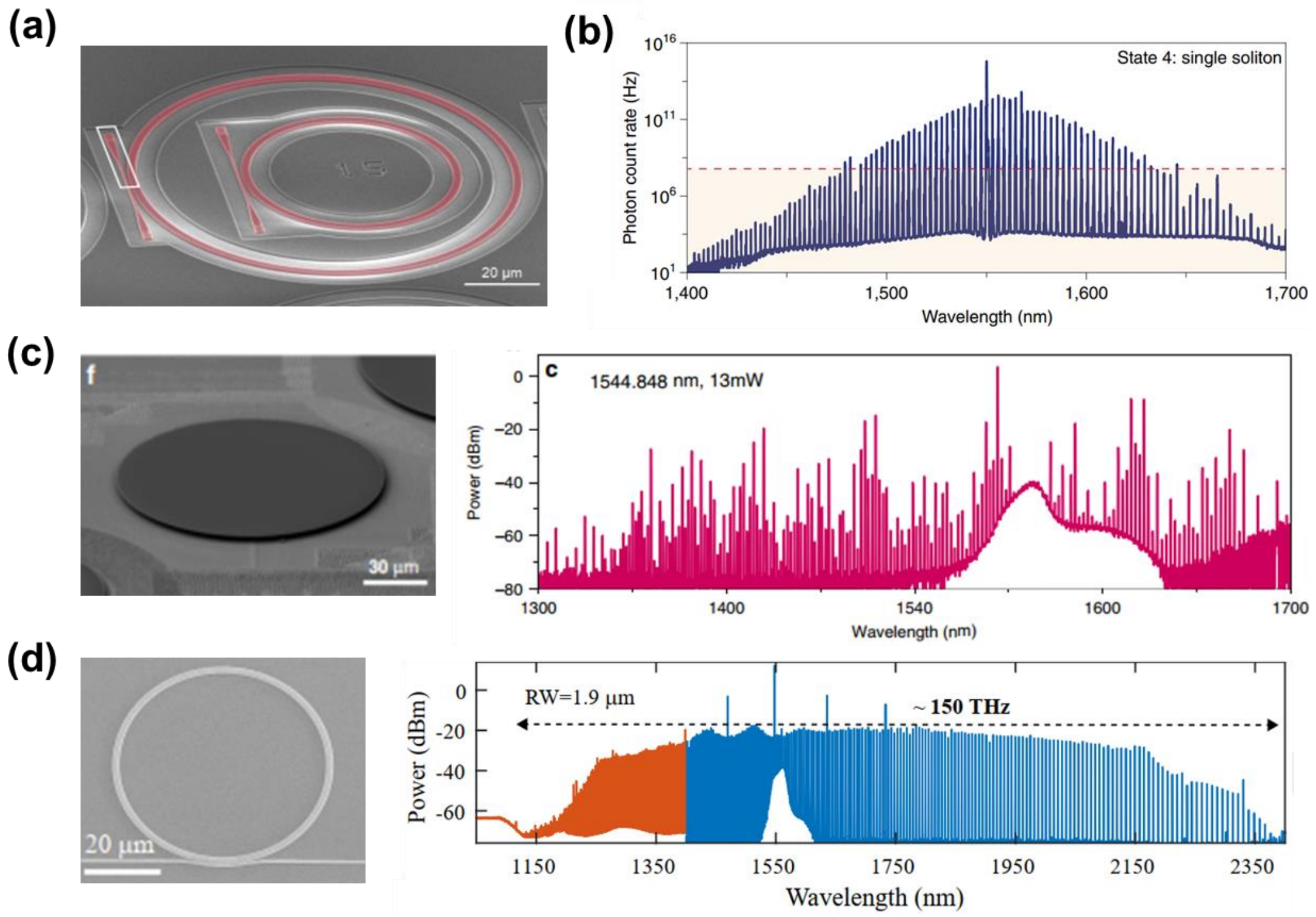

**Figure 11.** SiC-based micro-resonators for Kerr frequency comb generation. (a) A false-color SEM image of fabricated 4H-SiC micro-ring for micro-comb formation. Reprinted with the permission from M.A. Guidry *et al.,* Optica **7**, 1139 (2020). Copyright 2020 Optical Society of America[26]. (b) The single soliton state. Reprinted with the permission from M.A. Guidry *et al.,* Nat. Photonics **16**, 52 (2022). Copyright

2021 Springer Nature Limited[29]. (c) A micro-disk resonator (left) used for broadband Kerr frequency comb generations (right). Reprinted with the permission from C. Wang *et al.,* Light Sci. Appl. **10**, 139 (2021). Copyright C. Wang *et al.* 2021[30]. (d) A micro-ring resonator (left) for broadband comb generations (right), with the bandwidth spanning from the wavelength range of 1100 nm to 2400 nm[126]. Reprinted with the permission from L. Cai *et al.,* Photonics Res. **10**, 870 (2022). Copyright 2021 Optical Society of America.

M.A. Guidry *et al.*[26] demonstrated on-chip low-threshold (8.5±0.5 mW) OPO and micro-comb formation by leveraging micro-rings with $Q$ factors exceeding 1.1 million, as shown in Fig. 11(a). The authors determined a nonlinear refractive index for 4H-SiC of $n_2 = 6.9 \pm 1.1 \times 10^{-15}$ $cm^2$ $W^{-1}$ at 1550 nm by measuring the OPO power threshold in their devices. Going forward, lower frequency comb operation power to the levels that the integrated laser can reach (ten of mW)[127] needs to be achieved. Recently, a soliton micro-comb has been reported by the same group as shown in Fig. 11(b)[29]. At cryogenic temperature 4K, the stably generated SiC soliton micro-combs were used to investigate its quantum formation dynamics. Kerr frequency combs covering from 1300 to 1700 nm were demonstrated at a low input power of 13 mW using a dispersion engineered SiC micro-disk resonator[30], as can be seen in Fig. 11(c). For practical application, the bandwidth of a micro-comb should be extended to one octave or more in order to achieve *f-2f* self-referencing. An octave-spanning micro-comb was reported in the 4H-SiCOI platform as shown in Fig. 11(d)[126]. Although the observed octave-spanning comb state is pumped by high power near 120 mW and is not soliton, obtaining such a broadband comb represents a crucial step to realize f-2f self-referencing.

## 5. Optomechanical Properties

Aside from optical modes, resonant cavities with high-$Q$ factors can also support mechanical vibration modes, offering a promising system to study the interaction between electromagnetic radiation and mechanical motion[128]. Thus far the optomechanical properties have been extensively studied in various materials including silicon[129], silica[130], $Si_3N_4$[131], GaAs[132], and diamond[133]. Although SiC is rarely used in this field at present, its excellent optical and mechanical properties make SiC an excellent choice for optomechanical application.

The motivation for cavity optomechanics comes from the fundamental effects of quantum physics, gravity and ultra-sensitive sensing[134], quantum information processing[128,135]. High-performance optomechanical cavities are characterized by both strong-confined optical modes with high-$Q$ factors and high-$Q$ mechanical modes. The former depends on the optical performance of the material, as well as nano-fabrication process, and the latter largely depends on the intrinsic properties of the material. Large acoustic velocity and low material damping can facilitate mechanical resonance with high-frequency and high-$Q$ mechanical modes. Among diverse materials that are currently investigated, SiC exhibits outstanding mechanical properties with a bulk acoustic velocity of 12490 m $s^{-1}$ (in contrast, LN is 6541 m $s^{-1}$, silica is 5839 m $s^{-1}$). Recently, a theoretical calculation[22] suggested that SiC has lower intrinsic mechanical loss than other materials, with a frequency-quality product ($f_m \cdot Q_m$) as high as 6.4 × $10^{14}$ as compared to Si (3.9 × $10^{13}$) and diamond (3.7 × $10^{13}$). This high value can be ascribed to the ultra-low energy dissipation in crystalline SiC in Akhiezer regime[136], where the dissipation mechanism is mainly bound by anharmonic phonon-phonon scattering[125,137].

The most frequently used optomechanical resonators are micro-disk, especially double-layer micro-disk composed of a couple of micro-disks spaced by a thin film. The double-micro-disk resonator has ever been studied and proved to be an effective optomechanical resonator, providing strong coupling interaction between multi-optical and mechanical modes. It is assumed that the displacement changes the resonant frequency of the optical mode linearly[138]. The optomechanical coupling coefficient is described as $g_{om} \equiv d\omega/dx$, where $\omega$ is the resonance angular eigenfrequency and $x$ is the vertical mechanical displacement amplitude. For a WGMs resonator, $g = -\omega_0/r$, where $r$ is the radius of the resonator, $\omega_0$ is the resonance frequency. Numerous micro-resonators and micro-actuators based on 3C-, 4H- and 6H-SiC have been reported, confirming the excellent optical and mechanical performance of the SiC materials.

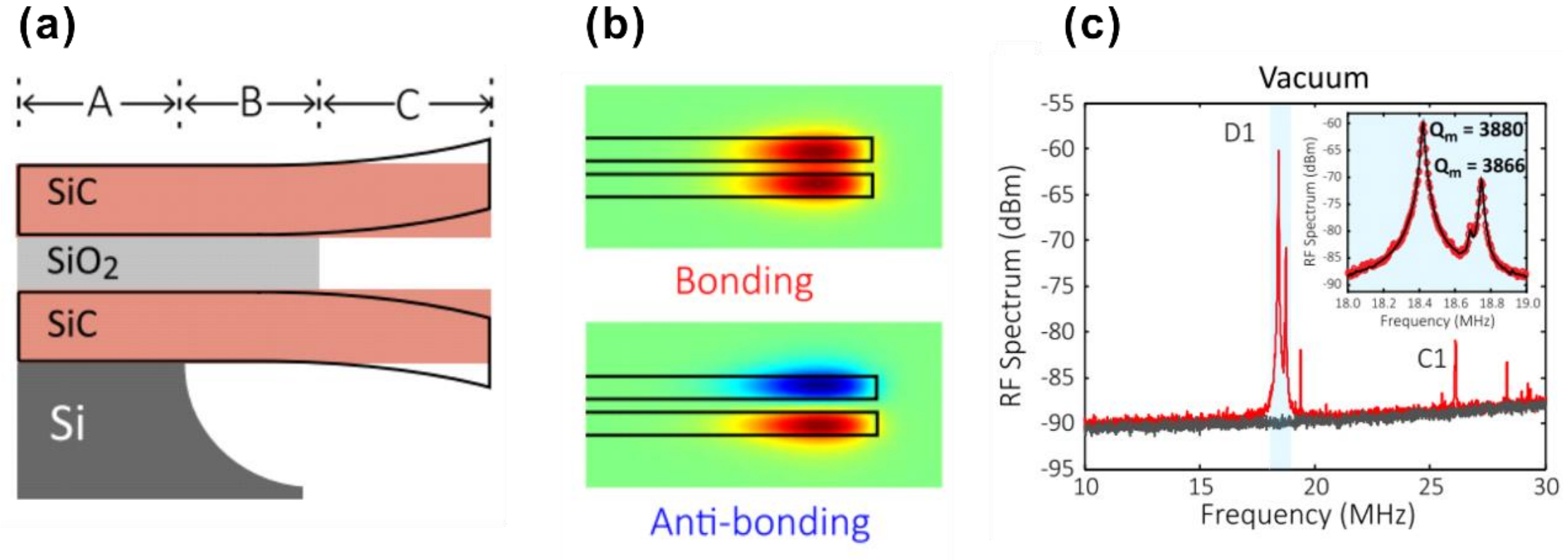

**Figure 12.** Silicon carbide double-micro-disk resonator. (a) Schematic view of a double-micro-disk resonator. (b) Numerical simulation of two types of fundamental optical mode profiles confined in double-micro-disk. The upper/bottom is bonding mode with the same/contrary phases (c) Optomechanical spectrum in vacuum show thermal-Brownian motion induced fundamental differential (D1) and common (C1) modes. Reprinted with the permission from X. Lu *et al.,* Opt. Lett. **44**, 4295 (2019). Copyright 2019 Optical Society of America[93].

Z. Wang *et al.* observed the intrinsic Brownian thermomechanical vibrations in a 3C-SiC micro-disk resonator[139]. Thanks to the excellent mechanical properties of SiC, ninth vibration modes were clearly distinguished by scanning spectromicroscopy measurement techniques, making it possible for coupling optical mode or quantum defect with these high-$Q$ mechanical modes. In 2015, X. Lu *et al.*[86] demonstrated SiC optomechanical resonator with high-$Q$ WGMs and high-$Q$ mechanical mode for the first time. The fabricated micro-disk with a radius of 2 μm showed a higher coefficient $|g_{om}|/2\pi \approx 80$ GHz nm$^{-1}$. The mechanical $Q$ value measured at 1.69 GHz was as high as 5.589× $10^3$, corresponding to $f_m \cdot Q_m$ product of 9.47 × $10^{12}$. The product is the highest value among any other WGMs optomechanical resonators fabricated on various materials, such as silica[130], $Si_3N_4$[140], GaAs[132]. The optomechanical resonator shown an optical $Q$ factor of $3.9 \times 10^4$, which needed to be further improved by optimizing the nano-fabrication process. Double-micro-disk resonators are considered to be more suitable cavities for optomechanical application, which can achieve a stronger optomechanical coupling than a single micro-disk. This has been realized on silica[141] and $Si_3N_4$[142] platforms. Most recently, X. Lu *et al.* firstly demonstrated the double-micro-disk resonator on 3C-SiC on Si platform[93]. Owing to the strong optomechanical interaction, the thermal-Brownian motion of mechanical modes[139] is directly observed by the continuous-wave lasers pumping. As shown in Fig. 12, the fundamental differential modes have a mechanical $Q$-factor > $3.8 \times 10^3$ in vacuum and the coupling coefficient is about $|g_{om}|/2\pi$ = 100 GHz $nm^{-1}$ due to the dependence on vertical dimension. The pair of micro-rings were separated by a nanoscale gap and supported by very thin spokes and a pedestal. By introducing the well-optimized thin spokes, the mechanical stiffness of resonators was reduced, thus increasing the sensitivity to the optical force between the rings.

Crystals with artificial periodic medium structure can simultaneously localize and manipulate the optical and mechanical modes, that is, a system simultaneously confining photons and phonons, which offers an opportunity to access novel optomechanical systems[143]. The common optomechanical crystals such as single-nanobeam and zipper photonics crystals have been demonstrated on SOI platform[144], $Si_3N_4$[145], and diamond[146,147]. Similar to the micro-disk resonators, the strength of interaction can also be quantified by the optomechanical coefficient $g_{om} \equiv d\omega_0/dx = \omega_0/L_{OM}$, where $L_{OM}$ is the effective coupling length over which the momentum of a photon is exchanged with the mechanical modes. According to the definition, smaller $L_{OM}$ implies a faster exchanging rate, and thus a stronger frequency response. Very recently, X. Lu *et al.*[99] demonstrated a zipper optomechanical crystal in 3C-SiC with strong coupling $g_{om}/2\pi$~10 GHz/nm and optical $Q$ factor of $2.8 \times 10^3$. The optical $Q$ of 4H-SiC-based nanobeam[32] is optimized as high as about $6 \times 10^5$, which may permit a much better optomechanical crystal. The high coupling efficiency optomechanical mechanics may coherently interact with the ground state of the spin qubits[31], combining with the significant nonlinearity of SiC. It holds a great promise for new exciting experiments on exploring mechanical manipulation approaches for integrated nonlinear and quantum photonics.

## 6. Integrated quantum photonics

SiC can host numerous optical optically-addressable spin defects working from visible (600–800 nm) to near-IR (850–1,600 nm) spectral range as schematically shown in Fig. 13(b). Many of them have been proven to have outstanding quantum properties such as bright single photon emission and long spin coherence time properties[148]. Unlike diamond, the CMOS compatibility of SiC allows for mass production of spintronic devices and meanwhile enables easy integration in photonic circuits. Therefore, SiC is an attractive material for exploring integrated quantum photonic circuits that involve desired spin-defects. In this chapter, we give an overview of color centers explored in SiC and the recent progress in the integration of these color centers with nano-photonic devices.

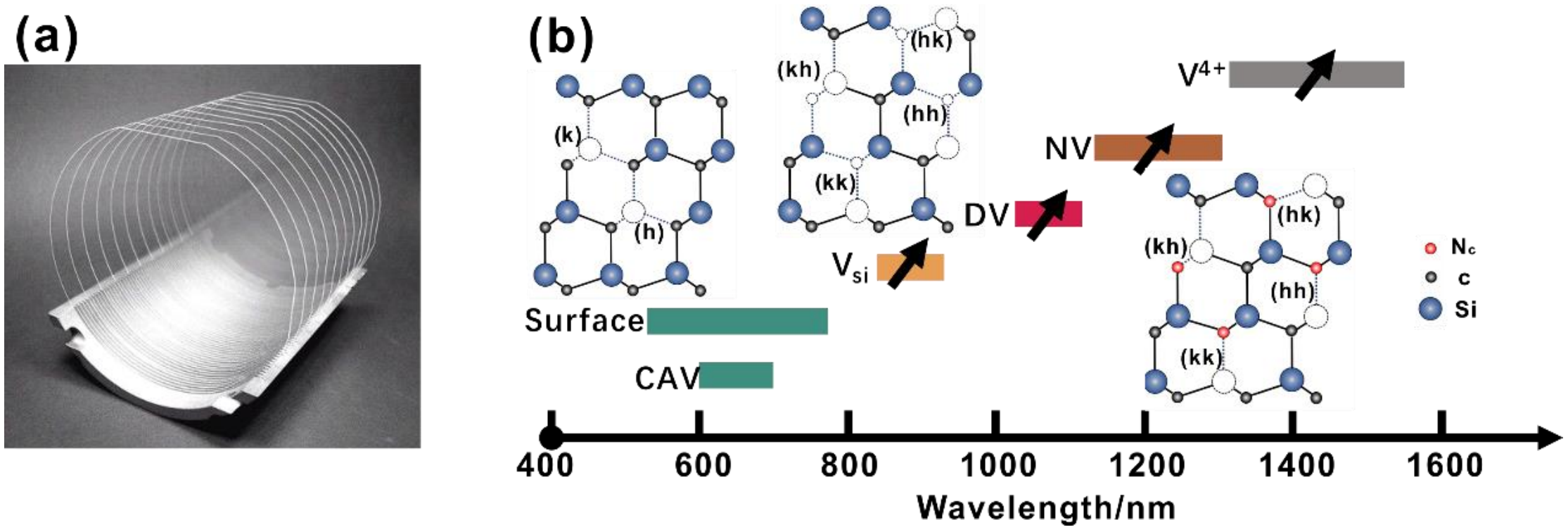

**Figure 13.** Wafer-scale high purity SiC and its color centers. (a) High pure commercial 4-inch SiC product. (b) Spectral map of various color centers in SiC which show single photon properties. The black arrow indicates that the spin center is reported to be coherently controlled. Insert figures show the defect locations of $V_{Si}$, DV, NV in 4H-SiC.

Quantum emitters have been observed in many systems including single molecules[149], atomically thin materials[150], quantum dots[151], color centers[152]. Among these, the color center system represented by the nitrogen-vacancy (NV) centers in diamond is considered to have more favorable quantum properties and has become the dominant platform for quantum applications[153]. Various color centers hosted in SiC also enjoy most of favorable properties of NV centers in diamond[154], significant progress has been made to explore the quantum properties of color centers in SiC[155-158]. More than 10 years ago, the abundant defects in the different SiC polytypes were discovered through photoluminescence and electron paramagnetic resonance measurements[155,156]. Recent researches mainly purposed to isolate the color center[46,159,160] and assessed its quantum properties more accurately[161,162], which is the fundamental issue in solid-state-based quantum technologies. The isolated defect shows single photon emitter properties and the hyperfine spin-related parameters has wide applications in magnetometry, quantum information and the exploration of fundamental physics. Optical detection magnetic resonance (ODMR) technology is widely used to coherently control the spin states and readout of the quantum characteristics of the defect center such as zero field splitting (ZFS) and spin coherence time[163,164]. 4H, 6H and 3C polytypes are commonly used in SiC color centers studies.

**Table 4.** The intrinsic quantum properties of color centers in 4H, 6H and 3C-SiC polytypes

| Material | color centers[(a)] | ZPLs[(b)] [nm] | Max. Counts for SPS [kcps] | ZFS[(c)] [MHz] | Spin | Fluorescence Lifetime[ns] | Spin coherence time T2 [ms] | DWF[(d)] [%] |
|---|---|---|---|---|---|---|---|---|
| 4H-SiC | $V_{Si}$ | 858-917 | 10 | 4, 70 | 3/2 | 5.5 -6.2 | 0.6-20 | 40 |
| | DV | 1078-1130 | 150 | 1224-1336 | 1 | 14 | 1 | 5 |
| | CAV | 648.7-676.5 | 2000 | /[(e)] | 1/2 | 1.8 | / | / |
| | NV | 1180-1240 | 17.4 | 1193-1331 | 1 | 2.1 | 0.017 | / |
| | $V^{4+}$ | 1279, 1335 | / | / | 1/2 | / | 0.0002 | <50 |
| 6H-SiC | $V_{Si}$ | 865-907 | / | 2.8-226 | 3/2 | / | / | / |
| | DV | 1093-1140 | / | 1236-1383 | 1 | / | / | / |
| | NV | 1278-1345 | / | 1240-1305 | 1 | / | / | / |
| | $V^{4+}$ | 1309-1389 | / | 16000-524000 | 1/2 | 11-108 | / | <45 |
| 3C-SiC | $V_{Si}$ | / | / | / | / | / | / | / |
| | DV | 1127 | 26 | 1300 | 1 | 18.7 | 0.9 | 7 |
| | NV | 1468 | / | 1303 | 1 | / | / | / |

[a] The names of various color centers are abbreviated as $V_{Si}$, silicon vacancy, DV, divacancy, CAV, carbon antisite-vacancy pair, NV, nitrogen vacancy pair, $V^{4+}$, neutral charge state vanadium defect, also labeled as $V^{(0)}$; [b] ZPL, zero-phonon line; [c] ZFS, zero-field splitting; [d] DWF, Debye-Waller factor; [e] '/' indicates no functionality reported, to the best of our knowledge.

Table 4 lists the quantum properties of different color centers reported in 4H-, 6H- and 3C-SiC. ZPL represents the zero-phonon line, which originates from the zero-phonon transition between energy levels. Unlike diamond, ZPL of color centers in SiC systems usually becomes visible at cryogenic temperature[46,159,160]. ZFS is the zero-field splitting measured by CW-ODMR test as mentioned earlier. DWF stands for Debye-Waller factor and can be calculated by computing the proportion of ZPL in the total PL spectrum. Spin coherence time is usually measured by pulse ODMR's spin echo. A key challenge for spin qubits system is to achieve a longer spin coherence time than the time for spin-based information storage and manipulation protocols[165]. Other quantum properties of the color centers such as spin-number, lifetime and maximum emission counts for single-defect are also listed. All these color centers are usually formed by substitutional impurities or lattice vacancies, such as defects of transition metal impurities and defects related to C, Si vacancies or substitution. Vacancy-related defects are usually caused by the irradiation of particles such as electrons[46,159,160], ions[47,166], and neutrons[167]. Due to different preparation conditions (such as epitaxy or sublimation method), high-purity SiC crystal products usually contain $V_{Si}$ vacancies. More complex DV and CAV defects need to be formed by annealing to reorganize the defect migration[46,159]. In addition, femtosecond laser processing can also form vacancy defects[168]. Regarding the preparation methods of various color centers, a review in 2020 has a very detailed summary[169].

To exploit the full potential of optical addressable spin defects in SiC, isolating a single color center with a suitable density is required. The first observed single-color emission in SiC is CAV in 4H polytype[46], where CAV represents for carbon anti-site-vacancy pair. By adjusting the electron irradiation dose and post annealing temperature, the density of CAV can be controlled from ensemble to single color center. The CAV has a bright room-temperature luminescence of around 700 nm, and the saturation emission of isolated CAV up to 2000 kcps is much higher than other color centers in SiC. Eight ZPLs are called AB lines, labelled as A1 = 648.7 nm, A2 = 651.8 nm, A3 =

665.1 nm, A4 = 668.5 nm, B1 = 671.7 nm, B2 = 673.0 nm, B3 = 675.2 nm, and B4 = 676.5 nm, respectively The AB lines generate from four unequal lattice positions and their corresponding excited state energy level splits. Although the first-principles calculations indicate that neutral charge state CAV can be manipulated[170], no ODMR has been tested experimentally. This emission has been associated with isolated single-photon sources. CAV was also studied in 3C-SiC as a possible single-photon emitter.

Aside from CAV color center, other important centers in SiC are negatively charged $V_{Si}$ and neutral DV. In 2010, it was reported that DV in 4H-SiC can perform spin coherent manipulation like NV in diamond at room temperature[164]. After a few years, DV and $V_{Si}$ were isolated and coherently controlled. $V_{Si}$ is a point defect formed by removing one Si atom in the lattice. It is found that the $V_{Si}$ centers are ubiquitous in 4H, 6H, and 15R-SiC, but rarely reported in 3C-SiC. As the vacancy locations which are shown in Fig. 13(b), silicon vacancy can occupy different sites in the 4H-SiC lattice and therefore $V_{Si}$ has different forms, including ZPL at 862 nm ($V_1$), 858.2 nm ($V_1'$) and 917 nm ($V_2$) for 4H-polytype. ODMR has not been observed for $V_1'$. For $V_2$ color center, ultra-long spin coherence time up to 20 ms has been reported, together with a saturation count of about 10 kcps. Regarding the optical properties of these color centers, the polarization direction of V1' is perpendicular to the *c*-axis, and the polarization directions of $V_1$ and $V_2$ are parallel to the *c*-axis[32,161]. Similarly, $V_{Si}$ has three unequal lattice positions in 6H-SiC, with the ZPL at 864.7 nm, 886.3 nm and 907 nm, corresponding to $V_1$ ($h$), $V_2$ ($k_1$), $V_3$ ($k_2$)[156,171], respectively. Recently, thanks to the stable and narrow linewidth emission facilitated by 4A2 symmetry of the excited and ground states of $V_{Si}$, high-precision spin coherent manipulation of $V_{Si}$ in SiC has been realized[161].

DV emissions have been reported in 3C-, 4H- and 6H-SiC, which are another promising candidate for spin qubits. In 4H-SiC, its associated ZPL are labeled as $PL_1$=1078 nm (*kh*), $PL_2$=1108 nm (*hk*), $PL_3$ =1131 nm (*hh*), and $PL_4$=1132 nm (*kk*), which correspond to the four non-equivalent sites[172] shown in Fig. 13(b). For the first time, DV single color center was isolated and coherently controlled by D.J. Christle *et al.* in an epitaxially high pure 4H-SiC wafer[159] with the DWF of 5.3%. The largest Hahn-echo spin coherent times of neutrally charged divacancy defects reached the millisecond range. The S = 1 spin number in the neutral charge state DV has a similar electronic structure to the NV center in diamond. $PL_1$-$PL_4$, $PL_5$-$PL_7$ have been assigned to DV configurations inside stacking faults. Most recently, the PL6 defect was isolated to single emission and demonstrated to coherent control[173]. Moreover, the single-photon saturation counting of $PL_6$ is up to 150 kcps[59], which is comparable with that of NV centers in diamond. Different from NV center in diamond and $V_{Si}$ center, the DV emission at the near telecom wavelengths can support a relatively low loss for fiber coupling, if desired. Combined with its high photon counts, the DV centers play an important role in long distance transmission for quantum communication network.

Similar to diamond, SiC is also able to host NV color centers with a total spin number $S = 1$. Most recently, coherent control of NV center in 4H-SiC was achieved at room temperature for the first time[59]. The maximum emission counts of single NV center is around 17.4 kcps, which can be equivalent to that of $V_{Si}$ and DV, and its NV emission is distributed in an infrared even communication band, with a emission wavelength in infrared spectral range. Other interesting defects in SiC emitting near telecom wavelengths can be achieved by implanting transition metal (TM) impurities, such as Mo[174], and V[175]. Some of these defects can be coherently controlled, as shown in Table 4.

Solid-state defects are capable to interface the solid-state spins with robust flying qubits (photons) that can transport the quantum information, which is the main factor for the interests in photonic applications. Notably, the creation of spin defects in nano-photonic structure renders on-chip integration convenience, thereby promising a considerable scalability for integrated quantum photonic circuits. From this prospective, a central challenge for this endeavor is to develop efficient photonic circuits to connect long-distance stationary qubit nodes[154]. Coupling color centers into high-$Q$ photonic cavities provides a possible route towards this goal as it allows for generation of indistinguishable photons and enhancement of spin-matter interaction by the Purcell effect, thus enabling the efficient generation, manipulation, and detection of color centers in photonic circuits[154]. This enhancement is quantified as the Purcell factor ($F$), which is given by

$$F = \xi \frac{3}{4\pi^2} \frac{Q}{V} (\frac{\lambda}{n})^3$$

where the parameter $\xi$ describes the overlap of the emitter's dipole with the cavity mode field. $V$ is the mode volume and typically expressed in units of cubic wavelength $(\lambda/n)^3$, $\lambda$ is the resonant wavelength of the cavity mode, and $n$ is the refractive index of the material. In order to maximize the Purcell enhancement effect, the cavity should have a good localization mode, high-$Q$ factor and small mode volume. Thus far, a variety of resonant cavities have been realized on SiC platforms (see Table 2 and Table 3). The successful fabrication of these devices is strictly connected to the capability of achieving an effective defect-cavity system in SiC, a nontrivial task for advanced photonic quantum technologies.

In 2014, G. Calusine *et al.* demonstrated the integration of ensemble Ky5 color centers into a PhC in 3C-SiC epitaxial film[47]. With the limitation of vertical sidewall, the authors achieved $Q$ up to $1.5 \times 10^3$ corresponding the color center emission collection efficiency enhancements to a factor of 10. In 2016, the same group demonstrated a further enhancement by a factor of 30, and the optical spin initialization rates from Ky5 were measured to be about 2-fold increase[176].1D nanobeam PhC cavities with $Q$ up to $7 \times 10^3$ was first demonstrated in homoepitaxially 4H-SiC using a dopant-selective etching method[64]. The $V_{Si}$ ensembles created by ion implantation were coupled to the cavities in the same work. To achieve a spectral overlap between the PhC resonant mode and the $V_{Si}$ ZPL, thinning the nanobeam film by RIE process and laser irradiating method were applied to tune the wavelengths of the cavity, thereby a 3-fold enhancement of the emission intensity was achieved. More efficient coupling of the PhCs and $V_{Si}$ was achieved by the same group in 2017[65], as shown in Fig. 14. The nanobeam PhCs with embedded $V_{Si}$ centers exhibit a $Q$ factor of $5.3 \times 10^3$, the $V_{Si}$ defects were created by $^{12}C$ ion implantation after the fabrication of the nanobeams, with ZPLs at 861.4 nm for $V_1$ and 916.5 nm for $V_2$, and the labeled $V_1'$ ZPL originates from the transition of an excited state with slightly higher energy (4.4 meV) than that of $V_1$ transition. As shown in Fig. 14(a), the cavity mode can be precisely redshifted to match the emission wavelength of the color center through a nitrogen gas condensation process.

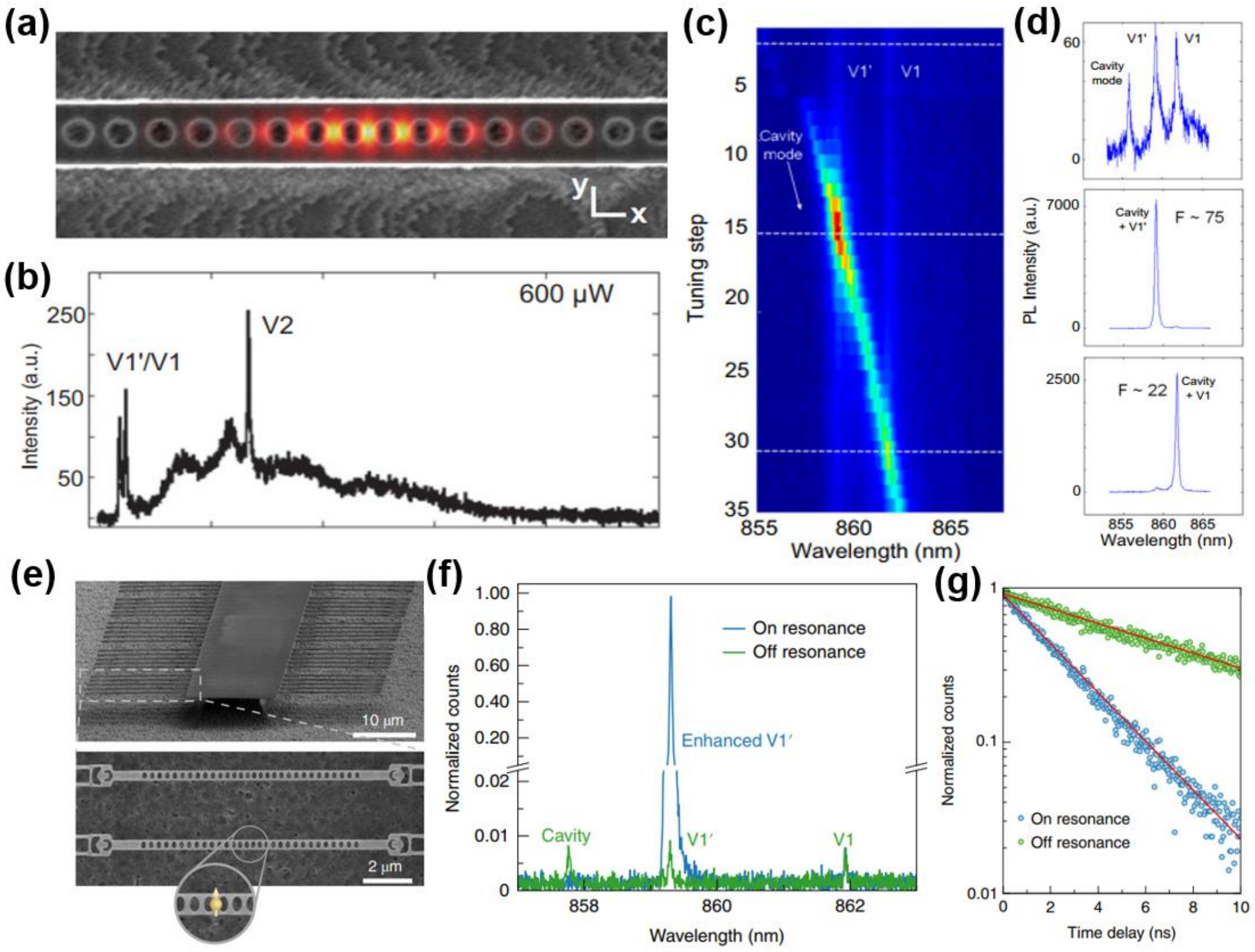

**Figure 14.** Characterization of SiC based quantum PhCs. The nanobeam PhC (a) with embedded $V_{Si}$ centers emission (b). (c) Stacked spectra map showing the interaction of the PhC cavity mode with the $V_1'$ and $V_1$ ZPLs as a function of gas condensation step. (d) Spectrum extracted from the mapping, the cavity mode far off-resonance (top), the cavity mode on resonance with $V_1'$ (middle) and $V_1$ (bottom). Reprinted with the permission from D.O. Bracher *et al.,* Proc. Natl. Acad. Sci. U. S. A. **114**, 4060 (2017). Copyright 2022 National Academy of Sciences[65].(e) SEM image of the 4H-SiCOI nanobeam cavity, integrated with coupled single spin qubit. (f) Spectrum with the cavity mode on-resonance and off-resonance. (g) A lifetime measurement with the cavity mode on-resonance and off-resonance. Reprinted with the permission from D.M. Lukin *et al.,* Nat. Photonics **14**, 330 (2019). Copyright 2019 Springer Nature Limited[32].

The top panel of Fig. 14(d) shows a spectrum of color centers when the cavity mode is far away from the two ZPLs. The middle and bottom panels of Fig. 14(d) show spectra when $V_1$ and $V_1'$ ZPLs were resonant with the cavity mode respectively. Comparison of the emission intensity before and after coupling yields the Purcell factor of 75 and 22, respectively. It was observed that the 9.4 ns off-resonant optical lifetime was reduced to 4.2 ns when the cavity was in resonance. According to the lifetime reduction effect, the Purcell enhancement was calculated to be 84. Although there is no specific DW factor value given in this work, the proportion of ZPL in the total PL spectrum reached an increase, which is expected to improve the entanglement rate between two spin-defects. The integration of a single neutral DV color center into a photonic crystal cavity was demonstrated[44], showing Purcell enhancement by a factor of 50 and a shortened on-resonant lifetime. The cavity-enhanced emitter led to a significant increase from ~5% to ~70-75% of the DW factor, and coherent control of the spin-cavity system was also achieved. This spin-cavity system can be of significance to the development of 4H-SiC as a platform for quantum information.

SiCOI provides a robust platform for integrated quamtum photonics for its mature and stable nano-fabrication techniques and the potential for wafer-scale applications. As disscussed in previous sections, due to the implantation damage, SiCOI prepared by the ion-cutting process has not been proved to have single-spin quantum properties. Recently, 3C and 4H-SiCOI can be prepared by wafer bonding and thinning method, High-$Q$ PhC cavities and micro-ring resonators can be fabricated by ICP etching processes[26,32,43,84,90]. Monolithic integration of single spin defects into photonic cavities has been commonly demonstrated in 4H-SiCOI platform[32], as indicated in Fig. 14(e). Using gas condensation tuning method, the emission from a single $V_1$ color center in the photonic resonator increased by about 120 times compared to the non-resonant emission. In the meantime, the life time of the photonic emission was reduced from 6.66 ns to 2.45 ns. The successful control of the optical properties of spin defects through the Purcell effect constitutes an important step towards realization of a high-efficient spin-photon interface in SiC. As an essential further step is to integrate these promising color centers in SiC in either monolithically or hybrid integrated photonic circuits, offering a pathway to develop the complex integrated photonic quantum circuits. Further investigation includes the realization of an efficient coupling with the nuclear spin states, and this allows entanglement distribution between stationary spin qubits mediated by photons via the Bell state measurements. Such a process, called spin-photon entanglement, is the heart of a quantum repeater proposed to increase the distance of quantum communication. It is expected that quantum photonic circuits could be successfully realized in SiC color centers, the same as what has been successfully performed for QDs in InAs/GaAs[177,178] and NV center in diamond[179,180].

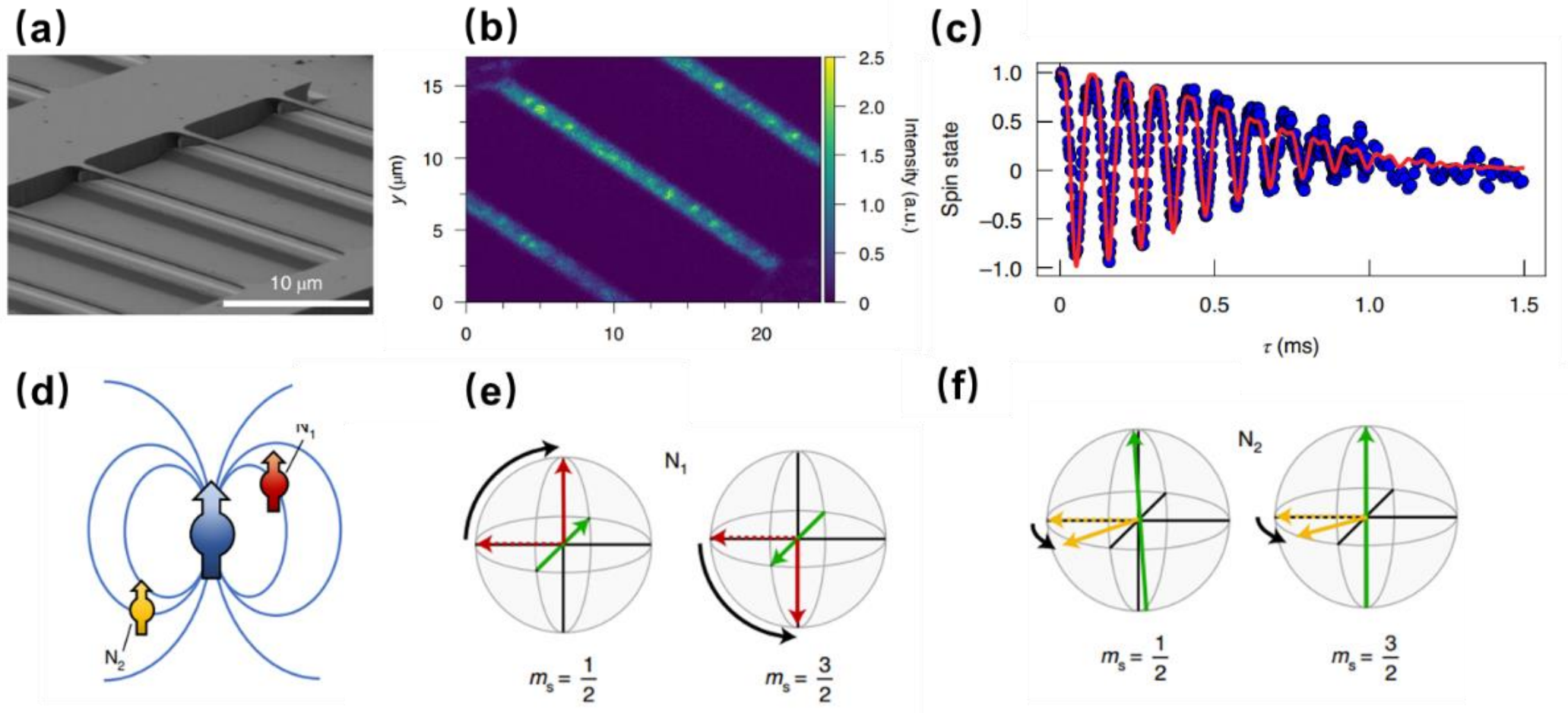

**Figure 15.** Properties of $V_2$ centers and control over nuclear spin qubits. (a) Schematic of a triangular cross-section waveguide with integrated electron and nuclear spins as fabricated in SiC. (b) Confocal fluorescence microscope image of the waveguides. Bright spots emission mainly from surface-related defects. (c) Spin-coherence of the implanted $V_{Si}$ in waveguides, with a spin echo time fitted to 0.84 $\pm$0.01 ms. (d) Schematic image of the coupled electron(blue) - nuclear spin (red and yellow, $N_1$ and $N_2$) triplet in the waveguide. Bloch-sphere representation of the controlled dynamics of the nuclear spins (e) $N_1$ and (f) $N_2$. Reprinted with the permission from C. Babin *et al.,* Nat. Mater. **21**, 67 (2022). Copyright 2021 Springer Nature Limited[39].

Although the single color center has been successfully created in nanofabricated SiC cavities, maintaining the optical and spin coherence properties of the qubits is regarded as a challenge. This is because that the nuclear-spin environment has been proven to be deteriorated by the implantation of scalable nano-photonics structures[44,181]. Recently, an array of $V_{Si}$ by implanting helium-ions into the SiC was first developed[39]. The low implantation energy of 6 keV was capable to reduce the crystal damage while introducing the color center with a suitable density. A $V_{Si}$ center was implanted into a nanofabricated waveguide, as shown in Fig. 15(a, b). Figure 15(c) shows the spin coherence time of the $V_{Si}$ in the nano-waveguides, indicating a value of 0.84 ± 0.01 ms, which is comparable to the bulk defect. This comparison demonstrated that the spin coherence time (about 1 ms) did not decrease for both the $V_{Si}$ center fabricated in bulks and those implanted in the waveguide. Furthermore, the authors used the high spin-optical coherences of $V_{Si}$ center in the waveguide to demonstrate the controlled operation of two nearby nuclear spins. Figure 15(d) schematically shows the electron spin of $V_{Si}$ coupled to $N_1$ and $N_2$ nuclear spins, forming a triplet. A spin sequence known as Carr-Purcell-Meiboom-Gill (CPMG)[182] is performed to observe the nuclei rotation induced modulation resonance. The Bloch spheres in and Fig. 15(e, f) show the electron spin dependent and independent rotation of $N_1$ and $N_2$ by applying to different CPMG pulses. The fidelity of the rotating operation was achieved as high as 98%, indicating that the waveguide coupled $V_{Si}$ centers can be efficiently utilized for controlling nuclear spins. Therefore, it has a great potential to become a leading competitor for integrated quantum application.

## 7. Outlook: opportunities and challenges

The ultimate goal of PICs is to transfer bulky free-space optics to miniaturized photonic chips, offering robust platforms to perform practical photonic technologies with enhanced complexity and stability. The recent successful demonstrations of both passive and active photonic devices in SiC, including low-loss optical waveguides, high-performance photonic cavities, EO light modulator and spin-based quantum light sources, are currently setting new state-of-the-art in the field of integrated photonics in terms of performance, flexibility and scalability. From this perspective, we have summarized various polytypes of SiC material for nano-photonic devices. Accompanied by the rapid development of passive photonic devices in bulk SiC and epi-3C-SiC, various preparation methods have been explored for developing SiCOI material platforms, including direct-wafer bonding, ion-cutting and thinning/polishing techniques. The potential and application of these platforms have been discussed and their strengths and shortcomings are also compared in this review paper. To achieve fully-fledged quantum applications, spin-based defects in both bulk SiC and thin-film SiC have also been discussed extensively in the paper.

Despite the remarkable process, research on SiC photonics is still in its infancy. The main challenge in this field is how to combine device engineering with the fundamental research. As the initial exploration, 3C-SiC on Si platform illustrates the possibility for photonic applications of SiC material. In terms of photonic devices platform, section 2.4.1 describes that hetero-epitaxial, bonding, and thinning methods may be a promising method for mass production of 3C-SiCOI. For 4H-SiC, angled etching and selective dopant approaches mentioned in sections 2.2 and 2.3 supply simple and efficient ways for the investigation of SiC color centers. The ion-cutting technique in section 2.4.2 can be used to prepare wafer-scale 4H-SiCOI with uniform thickness. Its photonic devices have also been verified and are expected to be commercialized. However, a large fluence of implanted ions will induce the deteriorating of the quantum property in SiC material, although the ensemble defect centers have also been discovered, a single-color center on this platform is still a serious challenge. Using SiC-$SiO_2$-SiC instead of SiCOI structure could provide another means to address the issue as SiC can support a higher post-annealing temperature, so that ion-induced lattice damages could be recovered. On the other hand, patterned implantation is also a way to make ion-cutting SiCOI micro-partially maintain quantum characteristics, which has already been proved in CMOS fabrication based on SOI platform[183]. Another critical issue is the related low thickness uniformity of the 4H-SiCOI thin-film prepared by the thinning and polishing techniques in section 2.4.3. This could be improved by employing industry-standard grinding and trimming equipments. It is also worth noticing that the waveguide loss fabricated on SiC (0.15 dB $cm^{-1}$) is far behind the other material platforms and remains a critical concern for the application of nonlinearity and large-scale photonic circuits where extremely low-loss propagation loss is required. As a summary, we conclude the following possible directions for SiC integrated photonics:

(1) Continuously improve the nano-fabrication techniques for reducing the propagation loss of waveguides. This can be achieved by employing additional chemo-mechanical polish lithography method as in the $LiNbO_3$ platform in order to suppress the surface roughness of the waveguide sidewalls[184]. The other way toward this optimization is to employ an optimal design where wider waveguide geometry is adapted so that the mode field is confined predominately in the center area.

(2) Develop proper methods to integrate SiC devices onto ultra-low loss photonic circuits in a hybrid fashion. Although SiC has shown a noteworthy potential to integrate high-performance nano-photonic devices, most polytypes of this material are prone to a relatively larger propagation loss which makes the improvement of the circuit complexity cumbersome. It is worth noticing that great efforts have been devoted to hybrid integration methods. In recent years, we have witnessed considerable progress in this field, and breakthroughs include the demonstration of deterministic self-assembled quantum dots-based single-photon sources with various low-loss photonic circuits have been realized, such as silicon, $Si_3N_4$ and so on[21]. In an effort to achieve complex and large-scale photonic circuits, if desired, integration of SiC-based passive and active devices developed thus far onto silicon-based low-loss photonic platforms that are amenable to CMOS-compatible fabrication technologies provide a feasible route.

(3) Recent studies have shown that absorption losses strongly depend on the orientation of the SiC crystallographic axis, *i.e.* the *c*-axis[26]. Currently, all polytypes of SiC with a single crystal phase have the *c*-axis perpendicular to the plane of the wafer. The best intrinsic absorption losses are obtained for SiC materials with the *c*-axis aligned in-plane. Therefore, specific single-crystal SiC wafers with the *c*-axis on the plane deserve more research work in the future. The wafer is particularly attractive for demonstrating high-performance EO light modulators. The present work shows that SiC materials with the *c*-axis perpendicular to the wafer plane can only be used to fabricate TM-mode-like EO modulators. Exploring SiC wafers with the *c*-axis in the plane facilitates the application of EO modulators based on TE-like modes. Also, this approach may generate new scattering losses, such as CMP and anisotropic etching issues, further exploration is still needed.

**Data availability statement**

The data that support the findings of this study are available from the corresponding author upon reasonable request.

**Acknowledgements**

This work was supported by Strategic Priority Research Program of the CAS (Grant No.XDC07030200), National Natural Science Foundation of China (No. 61874128, 11705262, 12074400 and 11905282), National Key R&D Program of China (2020YFB2008800)， Frontier Science Key Program of CAS (No. QYZDY-SSW-JSC032), Chinese-Austrian Cooperative R&D Project (No. GJHZ201950), Shanghai Science and Technology Innovation Action Plan Program (No. 17511106202, 20JC1416200), Program of Shanghai Academic Research Leader (19XD1404600), Shanghai Rising-Star Program (19QA1410600), Shanghai Sailing Program (No. 19YF1456200, 19YF1456400), K. C. Wong Education Foundation (GJTD-2019-11), Key Research Project of Zhejiang Laboratory under Grant 2021MD0AC01.